\documentclass[fleqn,usenatbib,onecolumn]{mnras}

\usepackage[T1]{fontenc}
\usepackage{ae,aecompl}

\usepackage{xcolor}
\usepackage{geometry}
\usepackage{natbib}
\usepackage{pifont}
\usepackage{dcolumn}
\newcolumntype{d}[1]{D{.}{.}{#1}}

\usepackage{graphicx}	
\usepackage{amsmath}	
\usepackage{amssymb}	
\usepackage{pdflscape} 
\usepackage{placeins}

\newcommand{\icm}{cm\textsuperscript{-1}}

\newcommand{\benz}{C$_{6}$H$_{6}$}

\newcommand{\cmark}{\ding{51}}
\newcommand{\xmark}{\ding{55}}

\title[Benzene on ice]{The vibrational properties of benzene on an ordered water ice surface}
\author[Clark \& Benoit]{Victoria H.J.\ Clark,$^{1}$\thanks{Contact email: \href{mailto:v.clark.17@ucl.ac.uk}{v.clark.17@ucl.ac.uk}}
and David M.\ Benoit,$^{2}$\thanks{Contact email: \href{mailto:d.benoit@hull.ac.uk}{d.benoit@hull.ac.uk}}\\
$^{1}$ Department of Physics and Astronomy, University College London, London, WC1E 6BT, UK \\
$^{2}$ E.A.\ Milne Centre for Astrophysics, University of Hull, Hull, HU6 7RX, UK} 

\date{Last updated 2019 July 19; in original form 2019 July 19}

\pubyear{2021}

\begin{document}
\label{firstpage}
\pagerange{\pageref{firstpage}--\pageref{lastpage}}
\maketitle

\begin{abstract}
We present a hybrid CCSD(T)+PBE-D3 approach to calculating the vibrational signatures for gas phase benzene and benzene adsorbed on an ordered water-ice surface. We compare the results of our method against experimentally recorded spectra and calculations performed using PBE-D3-only approaches (harmonic and anharmonic). Calculations use a proton ordered XIh water-ice surface consisting of 288 water molecules, and results are compared against experimental spectra recorded for an ASW ice surface. We show the importance of including a water ice surface into spectroscopic calculations, owing to the resulting differences in vibrational modes, frequencies and intensities of transitions seen in the IR spectrum. The overall intensity pattern shifts from a dominating $\nu_{11}$ band in the gas-phase to several high-intensity carriers for an IR spectrum of adsorbed benzene. When used for adsorbed benzene, the hybrid approach presented here achieves an RMSD for IR active modes of 21~\icm, compared to 72~\icm\ and 49~\icm\ for the anharmonic and harmonic PBE-D3 approaches, respectively. Our hybrid model for gaseous benzene also achieves the best results when compared to experiment, with an RMSD for IR active modes of 24~\icm, compared to 55~\icm\ and 31~\icm\ for the anharmonic and harmonic PBE-D3 approaches, respectively. To facilitate assignment, we generate and provide a correspondence graph between the normal modes of the gaseous and adsorbed benzene molecules. Finally, we calculate the frequency shifts, $\Delta\nu$, of adsorbed benzene relative to its gas phase to highlight the effects of surface interactions on vibrational bands and evaluate the suitability of our chosen dispersion-corrected density functional theory.

\end{abstract}




\section{Introduction}

Benzene, \benz, is an aromatic hydrocarbon which was first isolated and identified on Earth in 1825 by Michael Faraday \citep{faraday_1825}. Comprising of six carbon atoms bonded into a ring with a single hydrogen attached to each, benzene was first detected in the ISM by \citet{cernicharo_2001}. It is thought to be one of the most basic building block of polycyclic aromatic hydrocarbon (PAH) molecules and is a key ingredient for the synthesis of PAH molecules, if formed via a bottom-up approach \citep{kaiser_2015, salama_2008}. Along with its detection in the ISM, benzene has been detected and identified in a number of astronomical environments such as the planetary nebula of CRL618 \citep{cernicharo_2001}, the comae of comets and asteroids \citep{schuhmann_2019} and meteoritic chondrites \citep{delsemme_1975}, to name a few. Benzene is also thought to be a key ingredient of Titan's atmosphere \citep{waite_2007} and it has been recently detected in the atmosphere of Saturn \citep{Koskinen_2016}. Due to its importance, benzene is often included as a component of kinetic reaction models when investigating interstellar dust grains, as \citet{jones_2011} showed it could be formed through a barrierless reaction. Protoplanetry and protostellar disks are another important area in which benzene is being searched for \citep{woods_2007}, a logical next step as hydrocarbon molecules including PAHs have already been observed in these and similar environments \citep{visser_2007, geers_2006, acke_2004, cernicharo_2001, matsuura_2004, walsh_2014}. Complex organic molecules such as benzene are difficult to locate in protoplanetry disk due to their low abundance and complicated spectra \citep{walsh_2014}, therefore accurate binding energies and spectroscopic peak positioning are vital for this type of work. 

There have been many studies investigating the vibrational frequencies of gaseous benzene (eg., \cite{stephenson_1984, christiansen_1998, hagen_1983, szczepaniak_1972}). However, spectroscopic investigations of benzene on (water-)ice in the ISM have often focused on the vibrational frequency shifts of the OH bands of water \citep{michoulier_2020} owing to the many known phases of water ice \citep{salzmann_2019}, or the reactions of benzene \citep{noble_2020, sivaraman_2014} rather than the vibrations of the molecules on the surface.

Currently there are few studies investigating the vibrational frequencies of benzene on a water ice surface \citep{hagen_1983, dawes_2018} with many authors using ice ``nanocrystals'', containing up to eight water molecules as models (eg, \cite{augspurger_1992,tabor_2015,miliordos_2016}). However, it has been shown that results for both vibrational and binding/desorption energy calculations vary depending on the nature of the surface model (cluster or actual surface) \citep{silva_1994,thrower_2009} and using large surfaces is likely needed to accurately reproduce experiment \citep{ma_2012}. For this reason, and to further aid the detection and identification, benzene has recently been included in an experimental work investigating the desorption and infrared spectrum of small aromatic molecules in water-containing astrophysical ices \citep{salter_2021}.

It is known experimentally that the binding of benzene to water ice is likely to occur through the $\pi$-hydrogen bonding of benzene with the dangling H--O bonds of the ice surface (see for example \citet{silva_1994}). This binding mode (i.e. benzene acting as a hydrogen-bond acceptor) has also been confirmed by experimental and computational studies on benzene--water clusters \citep{gotch_1992, benoit_1998, suzuki_1992, engdahl_1985} and the present study builds on our previous publication which showed the feasibility of this type of molecular arrangement \citep{clark_2019}. Hence, adsorption of benzene on the surface of interstellar ice is a strong possibility and this paper presents a study of the expected infrared spectra that would be detected upon adsorption. 

Here, we explore from first principles the vibrational signature of benzene on an ideal high-dipole ice surface to generate a suitable reference theoretical spectrum that could guide observations. We use the basal plane surface of ferroelectric water ice XIh (henceforth referred to as ``ice'') as an ordered model for crystalline interstellar ice. The proton-ordered phase of crystalline ice Ih, ice XI or XIh (see ``\textit{Proton ordering and reactivity of ice}'', \citet{raza_2012}, PhD thesis), has been suggested to exist on Kuiper belt objects in the solar system \citep{McKinnon_2005} although it remains difficult to detect with certainty, but progress in this direction is being made \citep{fukazawa_2006, arakawa_2009, wojcik_2014}. Nevertheless, this thermodynamically stable phase of water ice below 72~K is computationally simpler to model due to its straightforward proton ordering and has also been used in other studies as a model for crystalline water in the ISM \citep{kayi_2011}.

The paper is organised as follows: in Sec.~\ref{sec:theory}, we present a fully-anharmonic technique to compute the predicted vibrational signature of benzene adsorbed on ice XIh from first principles, with a detailed description of our computational details in Sec.~\ref{sec:compdet}. We evaluate the agreement of our calculations with available experimental infrared laboratory data in Sec.~\ref{sec:results} and present our conclusions in Sec.~\ref{sec:conc}.

\section{Theory}
\label{sec:theory}
\subsection{Vibrational spectra modeling}
\label{sec:vibration_th}
 The anharmonic calculations are performed using the VCIPSI approach developed in \citet{Scribano_2008}. We represent the potential energy surface (PES), $V(q_1,\ldots,q_n)$, as a sum of one-mode terms, $V^{\mathrm{diag}}$, and two-modes terms, $V^{\mathrm{coupl.}}$, a pairwise approximation suggested by \citet{jung_1996}. Thus for a system composed of $N$ vibrational modes we have:
\begin{eqnarray}
V(\mathbf{q})=V(q_1,\ldots,q_n)=V^{(0)}+\sum_{i=1}^N V^{\mathrm{diag}}(q_i)+\sum_{i=1}^N \sum_{j>i}^N V^{\mathrm{coupl.}}(q_i,q_j)
\label{eq:pes}
\end{eqnarray}
where $q_i$ are the normal modes of the system, $V^{(0)}$ is the energy of the minimum. Note that throughout this paper, we use rectilinear normal modes as PES coordinates. The VCIPSI approach represents the vibrational wave function as a sum of $n$-mode vibrational states (configurations), which are in turn described as product of single-mode wave functions (modals), $\varphi_{i}^{(\mathbf{k})} (q_i)$. Thus: 
\begin{eqnarray}
\Psi_\mathbf{k} (\mathbf{q}) = \sum_\mathbf{k} C_\mathbf{k} \left[\prod_{j=1}^{n} \varphi_{j}^{(\mathbf{k})} (q_i)\right]
\end{eqnarray}
where $\mathbf{k}=\{k_1,k_2,\ldots,k_n\}$ specifies the excitation quanta of each single-mode wave function and $C_\mathbf{k}$ are the coefficients of each configuration. The modal product is optimised using a vibrational self-consistent field approach (VCSF) and the configuration coefficients are optimised using a selective vibrational configuration technique detailed in \citet{Scribano_2008}. In the present study, we follow closely the approach outlined in \citet{benoit_2015} for water on a Pt(111) surface. 

\subsection{Infrared intensities calculations and line averaging}
\label{sec:intensity_average}
The infrared transition intensity from $\nu_i=0$ to $\nu_i=1$ for each line, $I_i$, is computed using the diagonal (uncoupled) one-dimensional vibrational wave function and a global dipole surface computed at the the PBE-D3/MOLOPT-TZV2P level of theory (see Sec.~\ref{sec:compdet} for details). The strength of the fundamental transition is given by the integral:
 \begin{eqnarray}
 I_i=\frac{8\pi^3N_{Av}}{3hc}\nu_i \left|\left\langle\varphi_{i}^{\nu_i=0} (Q_i)\right|\vec{\mu}_i(Q_i)\left|\varphi_{i}^{\nu_i=1} (Q_i)\right\rangle\right|^2
\end{eqnarray}
 where $N_{Av}$ is Avogadro's number, $h$ is Planck's constant, $c$ the speed of light and $\nu_i$ is the uncoupled, diagonal, fundamental frequency for mode $i$. The diagonal dipole moment surface is computed using:
 \begin{eqnarray}
\vec{\mu}({\bf Q})=\vec{\mu}(0)+\sum_{i=1}^{n}\vec{\mu}_i(Q_i)
 \end{eqnarray}
 where $\vec{\mu}(0)$ is the dipole moment at equilibrium and $\vec{\mu}_i$ is the dipole moment along normal coordinate $Q_i$. In practice, we compute the dipole moment at each 1-D grid point and then interpolate the dipole moment to a finer-meshed representation using a cubic-spline algorithm for each dipole vector component. Note that one limitation of the scheme is that it cannot realistically model intensity borrowing or combination bands, as those inherently rely on coupled dipole surfaces. 

The position of degenerate bands that should be equivalent by symmetry (such as $E_{1u}$-bands, for example) are averaged using an intensity weighted scheme. This is necessary since the VSCF procedure ignores symmetry by constructions (independent modals) and can lead to non-equivalent transition frequencies for $E$-symmetry modes (although this is partly corrected by the VCIPSI approach). Our averaging scheme for two non-equivalent $E$-bands $\nu_a$ and $\nu_b$, each having intensity $I_a$ and $I_b$ is given by:
\begin{eqnarray}
\nu_\mathrm{avg.}=\frac{I_a \nu_a+I_b \nu_b}{I_a+I_b}
 \end{eqnarray}
This scheme is applied throughout, but for transparency all individual computed mode frequencies are shown in supplementary information (Tables \ref{tab:booktabs-gas} and \ref{tab:booktabs-ice}). 

\subsection{Potential energy surface for benzene on ice vibrational calculations}
\label{sec:vibpes}
In order to simplify the vibrational calculation, we use a partial Hessian technique (details in Sec.~\ref{sec:compdet}) where the ice surface is held fixed while the adsorbed benzene molecule is allowed to vibrate. This type of approach has already been validated for adsorbed molecules on metal surfaces \citep{benoit_2011}. However, in the present case, the mass separation between adsorbate and surface is not as large. As this can potentially cause artefacts in our calculations, we checked that normal modes obtained when allowing both benzene and the topmost ice layer to move were sufficiently localised on the benzene fragment. Clearly, a more elaborate treatment of ``second-order'' phenomena, such as intensity transfer and dissipative surface dynamics, should consider those but this beyond the scope of the present study.

The potential energy surface used for the anharmonic vibrational calculations is obtained using a hybrid approach that follows our earlier work for water on a Pt(111) surface \citep{benoit_2015}. The diagonal part of the PES in Equation \ref{eq:pes}, $V^{\mathrm{diag}}(q_i)$, is computed using a hybrid approach while the coupling term in Equation \ref{eq:pes}, $V^{\mathrm{coupl.}}(q_i,q_j)$, is computed at the periodic DFT (``low'') level of theory. Thus we have:
\begin{eqnarray}
V(\mathbf{q})=\sum_{i=1}^N V^{\mathrm{diag}}_\textrm{HYBRID}(q_i)+\sum_{i=1}^N \sum_{j>i}^N V^{\mathrm{coupl.}}_\textrm{low}(q_i,q_j)
\label{eq:vibpes}
\end{eqnarray}
where the $V^{(0)}$ term has been subtracted as it is not relevant for vibrational calculations. 
\begin{figure}
\centering
\resizebox{0.75\columnwidth}{!}{
 \includegraphics{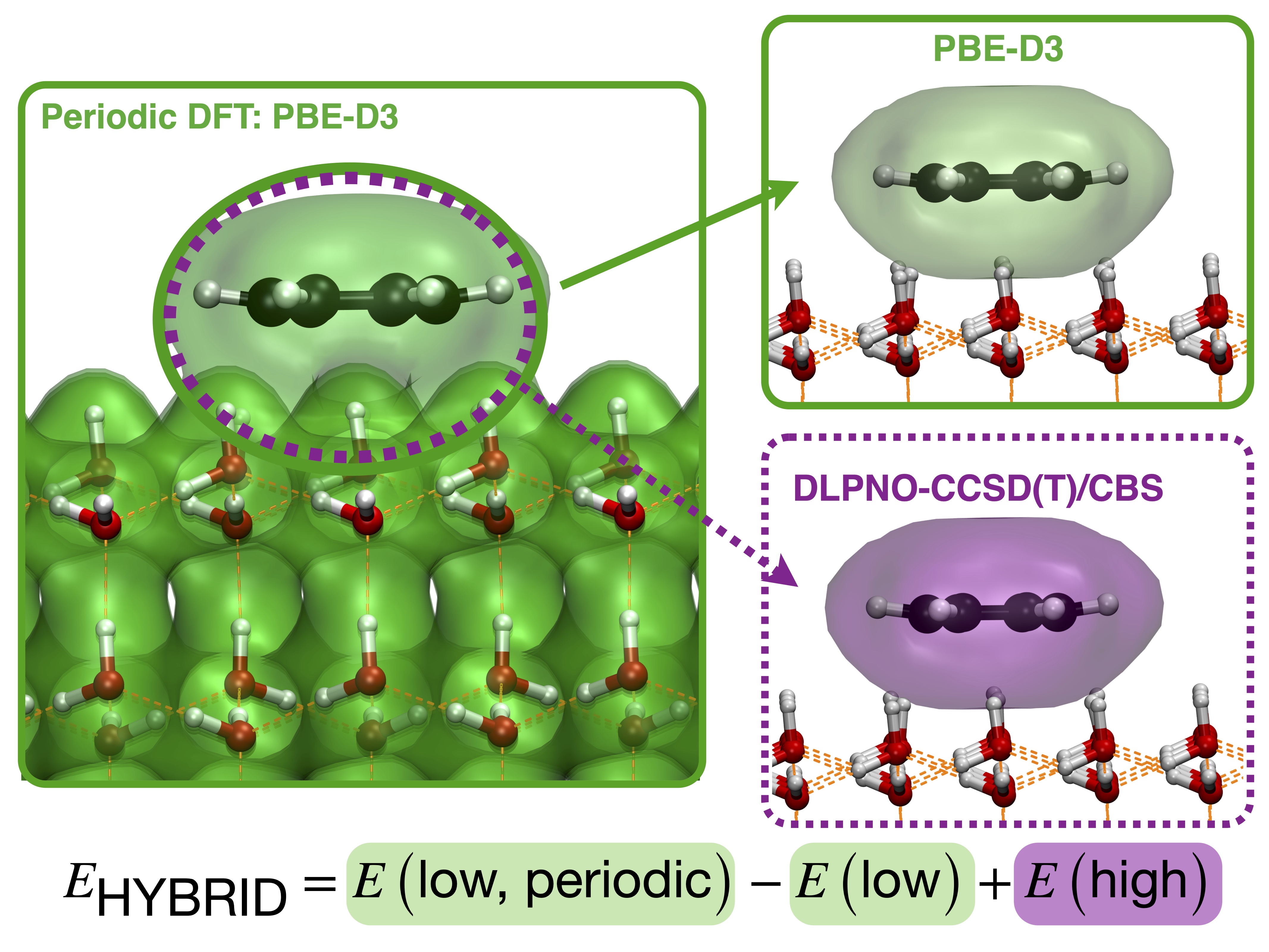}
}
\caption{Representation of ONIOM approach used to compute the hybrid potential. The energy is decomposed into three calculations: two periodic DFT calculations (full system and model, full line) and one DLPNO-CCSD(T) calculation (model, dashed line). The parts of the system include in each calculation are indicated using a green colour for PBE-D3 and a purple colour for DLPNO-CCSD(T).}
\label{fig:interactionsplit}
\end{figure}
In contrast to our previous study \citep{benoit_2015}, where we used an explicitly correlated F12 approach, the size of this system prevents us from using a high-level correlated method for the entire system. Therefore, we use an approach that mixes periodic density functional theory (DFT) with coupled-cluster singles and doubles with perturbative triples (CCSD(T)) calculations. The scheme, referred later on as DFT/CCSD(T), is a version of the ``our own n-layered integrated molecular orbital and molecular mechanics'' (ONIOM) approach developed by Morokuma and collaborators (see \citet{dapprich_1999}). This allows us to use a cost-effective yet high-quality approach and define $V^{\mathrm{diag}}_\textrm{HYBRID}(q_i)$ as:
\begin{eqnarray}
V^{\mathrm{diag}}_\textrm{HYBRID}(q_i)=V^{\mathrm{diag}}_\textrm{low}\left(q_i,\mathrm{\{Bz+Ice\}}\right)-V^{\mathrm{diag}}_\textrm{low}\left(q_i,\mathrm{\{Bz\}}\right)+V^{\mathrm{diag}}_\textrm{high}\left(q_i,\mathrm{\{Bz\}}\right)
\end{eqnarray}
where the notation $\mathrm{\{Bz+Ice\}}$, $\mathrm{\{Bz\}}$ indicates that the energy calculations are performed for benzene \emph{with} an ice surface or \emph{without}, respectively. The ``low'' level of theory is PBE-D3/MOLOPT-TZV2P and the ``high'' level of theory is DLPNO-CCSD(T)/CBS using non-augmented ano basis sets ($\mathrm{X}=3$ and $\mathrm{Y}=4$) and EC2 approach to speed up the PES generation. The energy decomposition scheme used is shown graphically in Figure \ref{fig:interactionsplit}. Further details of the method used are shown in Sec.~\ref{sec:compdet} below.

\section{Computational details}
\label{sec:compdet}
\subsection{Ice XIh model}
\label{sec:icedet}
Our model uses an ice surface generated from the optimised low-temperature proton-ordered coordinates from \citet{hirsch_2004}, obtained at the PW91/PW(350~Ry cutoff) level of theory (structure 1 in their original paper). In order to create the slab of ice XIh, the original unit cell ($a=4.49225$~\AA, $b=7.78080$~\AA~ and $c=7.33581$~\AA, $\alpha=\beta=\gamma=90^\circ$) was repeated to create a $6\times 3 \times 2$ supercell. In total there are 36 unit cells of XIh ice in the slab, totalling 288 H$_2$O molecules, with 4 double-layers of water molecules. The surface is chosen along the $c$ direction and thus the lattice constant is increased to $c=34.6716$~\AA~to accommodate the adsorption of a benzene molecule. In order to relax the surface energy, the positions of the surface atoms in the topmost water double-layer (72 H$_2$O molecules) are optimised, while the coordinates of the bottom three double-layers (216 H$_2$O molecules) of ice remains fixed. We use surface periodic boundary conditions in the $x, y$ ($a, b$) directions, while the $z$ ($c$) direction is treated non-periodically.

\subsection{DFT calculations}
All PBE-D3 calculations were performed using the Gaussian plane waves (GPW) method implemented in the QUICKSTEP module \citep{goedecker_1996, vandevondele_2005, perdew_1996} 
of CP2K (v4.1 and v6.1) \citep{vandevondele_2007, naumkin_1995, chergui_1996, goedecker_1996, lippert_1997}. The valence electrons of all atoms are described using a TZV2P-MOLOPT-GTH basis set \citep{dunning_1971, vandevondele_2007, krack_2005}. 
The core electrons are represented using Goedecker-Teter-Hutter (GTH) pseudo potentials \citep{krack_2005} along with an auxiliary plane-wave cutoff of 300~Ry. XY periodicity was used for all calculations with an analytical Poisson solver for the electrostatic energy. The wave function convergence was set to 1.E-7~a.u.\ for all calculations. During geometry optimisations, the convergence criterion for the maximum force of element was set to 1.E-5~bohr$^{\textrm{-1}}$~Hartree except for the geometry optimisation of gaseous benzene, when a maximum force of 1.E-6~bohr$^{\textrm{-1}}$~Hartree was required. The exchange-correlation functional is that derived by Perdew, Burke and Ernzerhof (PBE) \citep{perdew_1996} and we account for dispersion interactions using the DFT-D3 correction scheme of Grimme et al.\ based on a damped atom-pairwise potential and three-body $C_9$ corrections \citep{wishnow_1996}.
While the coordinates of the benzene atoms remained unconstrained in all calculations, the coordinates of all ice molecules were constrained in the vibrational analysis (288 molecules), while the top layer of ice was left unconstrained in the geometry optimisations (only constraining the lower three layers, corresponding to 216 H$_{2}$O molecules, as described in section \ref{sec:icedet}). The geometry of the adsorbed benzene and selected structural parameters were reported already in \citep{clark_2019}, for full transparency we have also included a coordinate file as supplementary information. 

\subsection{DLPNO-CCSD(T)/EC2-CBS calculations}
\label{sec:dlpnodet}
The complete basis set extrapolation of the domain-based local-pair natural orbital coupled-cluster singles doubles and perturbative triples (DLPNO-CCSD(T)) energy was performed using the ORCA 3.0.3 suite of programs. Note that this particular version of ORCA, the (T) implementation uses a ``semi-canonical'' approximation to compute the perturbative triples correction, also known as T0 correction. The expression used for the EC2-CBS extrapolation, due to \citet{jurecka_2006} is:
\begin{eqnarray}
E(\textrm{DLPNO-CCSD(T)/EC2-CBS(X,Y)}) &\approx& E(\mathrm{SCF;Y}) + E(\mathrm{DLPNO-CCSD(T);X})\nonumber\\
{}&& + E(\mathrm{MP2;}\infty) - E(\mathrm{MP2;X})
\end{eqnarray}
Where $\mathrm{X}$ and $\mathrm{Y}$ are the cardinal number of each basis set used. Here we use the ano-pVTZ basis set together with the ano-pVQZ basis set \citep{neese_2011} and thus $\mathrm{X}=3$ and $\mathrm{Y}=4$, respectively. 

The MP2 energy is extrapolated using:
\begin{eqnarray}
E(\mathrm{MP2;}\infty)=\frac{\mathrm{X}^\beta\cdot E(\mathrm{MP2,X})-\mathrm{Y}^\beta\cdot E(\mathrm{MP2,Y})}{\mathrm{X}^\beta -\mathrm{Y}^\beta}
\end{eqnarray}
where $\beta=2.41$, value optimised by \citet{neese_2009, neese_2011}.

In order to accelerate the calculations, we used both the RI-JK approximation \citep{weigend_2008} (with a cc-pVQZ/JK auxiliary basis set \citep{dunning_1989}) and the RI-MP2 approach \citep{weigend_1997} (with a aug-cc-pVQZ/C auxiliary basis set \citep{kendall_1992}). All calculations used the verytightscf convergence criterion of ORCA \citep{neese_2012}. All DLPNO-CCSD(T) calculations used the default ORCA criteria for the PNO generation (NormalPNO).

\subsection{Vibrational calculations}
We obtain the harmonic normal modes, $q_i$, by diagonalising a mass-weighted partial Hessian matrix computed in CP2K. Note that all surface atoms are kept frozen during the Hessian matrix calculation.

The potential energy surface computation is then driven by the \href{pvscf.org}{\textsc{Pvscf}} program \citep{benoit_2011} that is developed in our laboratory. We use the DFT energy values computed for 16 points along each each rectilinear vibrational coordinate to interpolate a finer-meshed representation of the diagonal terms (1-D) of the PES using a cubic-spline algorithm. The starting and finishing points of each coordinate scan in $q$-space are optimised to ensure that the curve supports at least eight bound states. Each mode--mode coupling term is computed for $16\times 16$ regularly spaced grid points and then interpolated on a finer mesh using a bicubic interpolation \citep{akima_1996}. These points form a discretised version of the PES of the system, $V({\bf Q})$, as formulated in Eq.~\eqref{eq:pes}. 

In order to provide high-accuracy surfaces, the 1-D energy points are re-computed using the DLPNO-CCSD(T)/EC2-CBS(ano-3,4) approach described in section \ref{sec:dlpnodet} for the benzene only and a hybrid ONIOM model for benzene on ice described in section \ref{sec:vibpes}. Note that this scheme only contains benzene in the ONIOM layer (i.e.\ no water molecules are included) and thus does not require diffuse (aug-) functions in the basis.

All the 1-D VSCF equations are solved using the FGH method \citep{marston_1989, balint-kurti_1991}. We compute the correlation corrections using the VCIPSI method and explore a VCI basis made of all one-, two-, three- and four-mode excitations (VCISDTQ) up to seven excitation quanta ($n_{\mathrm{max}}=7$). In particular, we note that in order to accurately model the C--H stretching bands, it is necessary to include modal excitations of up to 4 modes simultaneously, even with a simple mode--mode coupling. The convergence of these vibrational bands with the excitation level is shown in Table \ref{tbl:convergence} in the supplementary material. The iterative VCI matrices are diagonalised using our own implementation of the Davidson algorithm. Note that the vibrational calculation is only performed for the $\Gamma$-point of the Brillouin zone \citep{keceli_2010,ulusoy_2011}. 

All anharmonic transition intensities are computed using a diagonal representation of the dipole surface computed at the PBE-D3/MOLOPT-TZV2P level of theory along with the 1-D vibrational wave functions (i.e.\ without including effects from vibrational correlation).
 
 \section{Results and Discussion}
 \label{sec:results}
As a highly-symmetric molecule, benzene is a challenging system due to a number of resonances, high symmetry band groups and vibrational modes that combine both bending and stretching motions. This combination of factors stretches the capabilities of the simple harmonic approximation, likely leading to a poor overall mode description in a rectilinear coordinate system and inaccurate resonance description. 

However symmetry can easily be included in that model and offers a strong guiding principle for attributing vibrational modes to observed bands. When benzene is placed on an ice surface (even one with a symmetry that resembles that of benzene, like the XIh ice surface), most of the symmetry-based guiding principles are weakened (or even absent) thus making mode attribution extremely difficult. 

In this context, we compare the performance of the harmonic approximation to that of an accurate method that naturally accounts for vibrational resonances, namely the vibrational configuration interaction on a basis of vibrational self-consistent modals (VSCF/VCIPSI). The accuracy of the approach has already been shown for many molecules on surfaces (see for example \citet{chulkov_2013} or \citet{benoit_2015}). 

\begin{figure}
\centering
\resizebox{0.75\columnwidth}{!}{
 \includegraphics{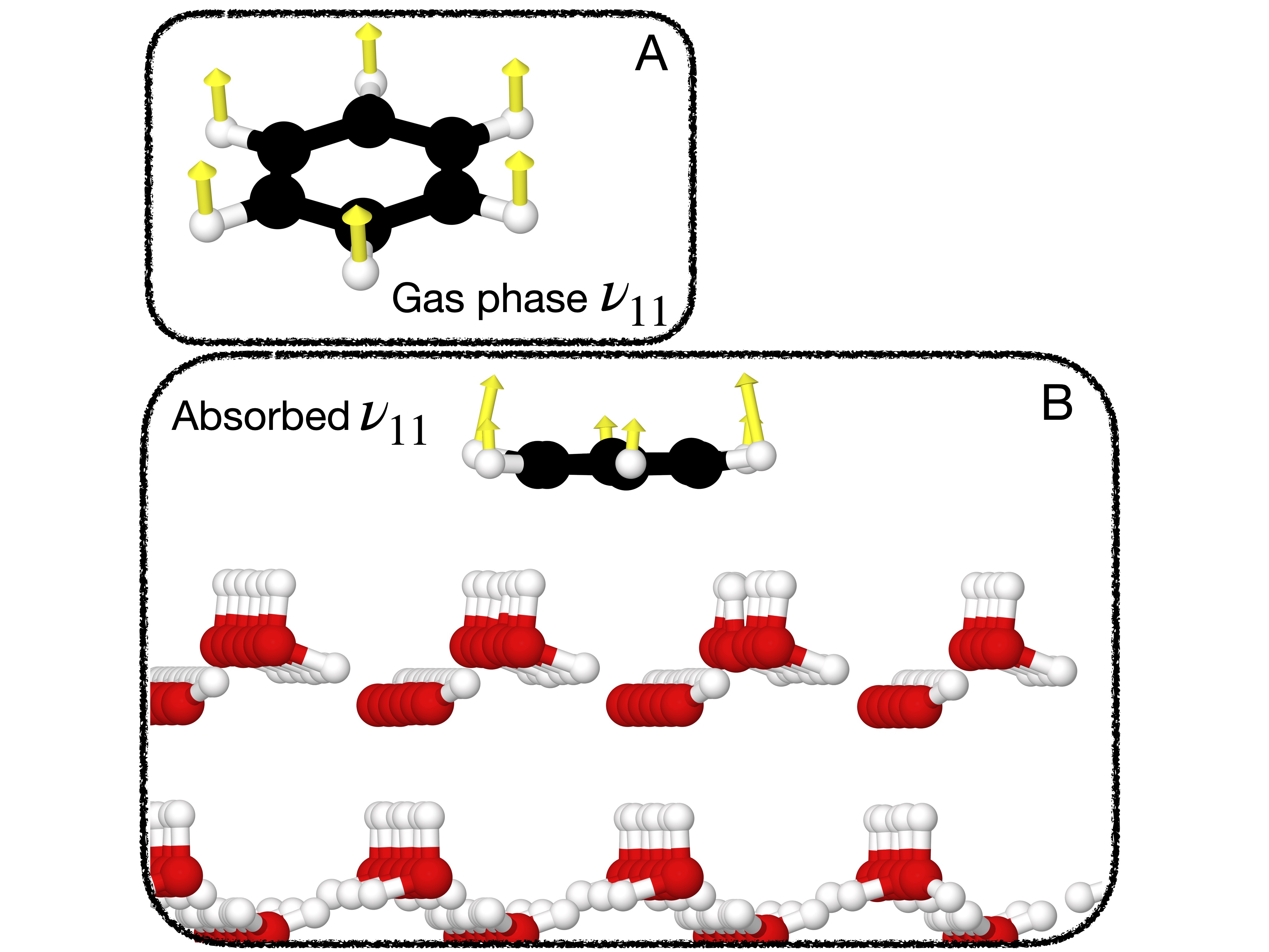}
}
\caption{Vibrational mode corresponding to out-of-plane band $\nu_{11}$ for benzene in the gas phase (Box A) and benzene adsorbed on a highly-ordered basal plane surface of ferroelectric water ice XIh (Box B). Note the difference between the symmetric appearance of the mode in the gas phase while the adsorbed version of this mode is noticeably asymmetric.}
\label{fig:bzmodes}
\end{figure}

\subsection{Gaseous benzene}
In this subsection, we compute the vibrational spectrum of gaseous benzene as a benchmark to establish the effect of surface interaction on its vibrational signature. We note that gas-phase benzene has already been the subject of many studies both experimental and theoretical (See references in Table \ref{tab:booktabs-gas} for specific experimental lines, and \cite{christiansen_1998, senent_2002} for examples of theoretical work), but our aim here is simply to correlate our adsorbed spectra (see Section \ref{sec:Bzicevib}) with results for a non-adsorbed system, obtained using the same approach. 

\begin{table}
 \centering
 \caption{ Infrared active vibrational frequencies in \icm\ of isolated gas-phase benzene. All degenerate computed bands are intensity averaged (see Sec.~\ref{sec:intensity_average} for details).}
 \begin{tabular}{@{} ccccd{4.3}c @{}} 
 \hline
 \hline
& Harmonic & \multicolumn{2}{c}{Anharmonic} & \\
\cline{3-4}
 Band [${\mathcal D}_{6{\rm h}}$ label] & PBE-D3 & PBE-D3 & CCSD(T)+PBE-D3 & \multicolumn{1}{c}{\textrm{Exp.}} & Ref.\\
 \hline
11 [A$_{\rm 2u}$]&	663&	694	&718& 673.975 & \citet{hollenstein_1990}\\
18 [E1u]& 	1034&	1026&	1051 &1038.267 &\citet{pliva_1984} \\
19 [E1u]& 	1465&	1443&	1490&1483.985 & \citet{pliva_1983} \\
20 [E1u]& 	3105&	2949&	3059&3047.908 & \citet{pliva_1982} \\
\hline
RMSD&31	&55	&24&\textrm{---} &\\
\hline
\hline
 \end{tabular}
 \label{tab:IRfreqs}
\end{table}

Our results for the infrared active modes are shown in Table \ref{tab:IRfreqs}. For each band, we report results obtained using a harmonic model (PBE-D3 surface), our anharmonic model (using two potential energy surfaces) and the latest experimental results. As is known from group theory, the infrared spectrum of benzene is surprisingly sparse (see Table \ref{tab:IRfreqs}), with the ${\mathcal D}_{6{\rm h}}$ point group of benzene allowing only 4 active vibrational bands (one A$_{\rm 2u}$ band: $\nu_{11}$, and three E$_{\rm 1u}$ bands: $\nu_{18}$, $\nu_{19}$ and $\nu_{20}$). We observe that those bands are qualitatively well described by all models, although we note that the harmonic model overestimates $\nu_{20}$ by over 50~\icm. Instead, our anharmonic approach leads to an excessive lowering of this band by $\sim$100~\icm\ relative to experiment for PBE-D3. This functional has already been shown to describe bends better than it does stretches (see \citet{respondek_2009} for further details). The hybrid PES leads to a welcome redshifting of $\nu_{20}$ to 3059~{\icm} in close agreement with the experimental value of \citet{pliva_1982}. 

However, the bands corresponding to modes that contain a bending component ($\nu_{11}$, see also Figure \ref{fig:bzmodes}, and $\nu_{18}$ to some extent) are poorly described by the hybrid approach and overestimate the experimental values (by 44~{\icm} and 13~{\icm}, respectively). This outcome is not unexpected as the vibrational self-consistent field (VSCF) theory ignores symmetry relations and uses a rectilinear coordinate system which typically disadvantages bends over stretches\footnote{Other coordinate systems are available in \textsc{Pvscf} but are untested for molecules on surfaces, see \url{https://pvscf.org} for further information.}. Indeed, bends involve angular motions which require a curvilinear coordinate system and cannot be described adequately with (recti)linear vectors. This has already been explored for methanol, for example, and clearly showed some of the shortcomings of the rectilinear approach for torsional modes, e.g., \citet{scribano_2010}. For benzene, the limitations of this coordinate system means that some of the complex bend-stretch bands (e.g.\ $\nu_{11}$ and $\nu_{18}$) are disadvantaged by our present approach and likely appear in the incorrect order.

Another consequence of the shortcomings of the rectilinear approach, combined with a limited 2-mode interactions PES, is shown for $\nu_{20}$, where our approach predicts a doublet instead of the single band of E$_{\rm 1u}$ symmetry seen experimentally (see Fig.~\ref{fig:ice-gas-comparison} and Table~\ref{tab:booktabs-gas}). Indeed, the splitting between the two modes (28 and 29) that make up $\nu_{20}$ is only 3~\icm\ at the harmonic level but further increases for the anharmonic calculation on both PBE-D3 and hybrid surfaces (16~\icm\ and 24~\icm, respectively). By de-coupling the C--H stretches from the rest of the modes, we can estimate that the exaggerated rectilinear stretching--bending coupling terms contribute as much as $10-15$~\icm\ and that the remaining differences are likely down to lack of a symmetrical exploration of the PES for those two equivalent modes.

Nevertheless, the root-mean-square deviation of the anharmonic approach on our hybrid surface is small (24~{\icm}) and improves on both harmonic and anharmonic approaches for PBE-D3 (31~{\icm} and 55~{\icm}, respectively).

The Raman-active bands provide further comparison points with experiment and our results for those modes are shown in Table \ref{tab:Ramfreqs}. Here again, we provide results obtained for the harmonic model (PBE-D3 surface), our anharmonic model (using two potential energy surfaces) and the latest experimental results. The ${\mathcal D}_{6{\rm h}}$ point group of benzene allows more bands for this type of spectroscopy, leading to 7 active vibrational bands (two A$_{\rm 1g}$ bands: $\nu_{1}$ and $\nu_{2}$, one E$_{\rm 1g}$ band: $\nu_{10}$, and four E$_{\rm 1g}$ bands: $\nu_{6}$, $\nu_{9}$, $\nu_{8}$ and $\nu_{7}$). 
\begin{table}
 \centering
 \caption{Raman active vibrational frequencies in \icm\ of isolated gas-phase benzene. All degenerate computed bands are intensity averaged (see Sec.~\ref{sec:intensity_average} for details).}
 \begin{tabular}{ @{}cccc d{4.4} c @{}} 
 \hline
 \hline
 & Harmonic & \multicolumn{2}{c}{Anharmonic} & \\
\cline{3-4}
Band [${\mathcal D}_{6{\rm h}}$ label] & PBE-D3 & PBE-D3 & CCSD(T)+PBE-D3 & \multicolumn{1}{c}{\textrm{Exp.}} & Ref. \\
 \hline
6 [E$_{\rm 1g}$]& 603&601 &609& 608.13 & \citet{hollinger_1978} \\
10 [E$_{\rm 1g}$]& 834&846 &879& 847.1062 & \citet{pliva_1996} \\
1 [A$_{\rm 1g}$]& 991 & 984&1000&993.071 & \citet{jensen_1979} \\
9 [E$_{\rm 1g}$]& 1162&1154 &1184&1177.776 & \citet{hollinger_1978} \\
8 [E$_{\rm 1g}$]& 1591&1566 &1620&1600.976 & \citet{pliva_1987} \\
7 [E$_{\rm 1g}$]& 3091&2936 &3041&3056.6 & \citet{hollinger_1978}\\
2 [A$_{\rm 1g}$]& 3115&2982 &3081&3073.942 & \citet{jensen_1979}\\
\hline
RMSD&22	&60	&16&\textrm{---} &\\
\hline
\hline
 \end{tabular}
 \label{tab:Ramfreqs}
\end{table}

We see, here too, that all Raman bands are qualitatively well reproduced by all models. The harmonic approach overestimates the high-frequency modes ($\nu_{7}$ by 34~\icm\ and $\nu_{2}$ by 41~\icm), while the anharmonic PBE-D3 results strongly underestimate the two same bands ($\nu_{7}$ by 121~\icm\ and $\nu_{2}$ by 92~\icm). The prediction of most Raman active bands is dramatically improved by the hybrid PES, which corrects both $\nu_{7}$ and $\nu_{2}$ to within 15~\icm\ and 7~\icm\ of the experimental data, respectively. One notable exception is the E$_{\rm 1g}$ band ($\nu_{10}$), which was fortuitously well described by both the harmonic and anharmonic treatment at the PBE-D3 level of theory but is overestimated by nearly 32~\icm\ in the hybrid model. Here rectilinear coordinates likely couple C--H stretching modes into this out-of-plane bending band and fortuitously favour the PBE-D3 anharmonic calculation, leading to an apparent good agreement for this band. 

Nevertheless, our hybrid approach leads to a very good overall agreement with experimental data (root-mean-square deviation of 16~\icm), compared to the PBE-D3 surface (22~\icm\ for the harmonic model and 60~\icm\ for the anharmonic model). 
This agreement is in line with the expected performance of our model (see \citet{scribano_2010} for a more detailed discussion) and further accuracy increase would require a more elaborate mode--mode interaction potential (3-modes or more) and/or a non-rectilinear coordinate system. 

\subsection{Benzene on an ice surface}
\label{sec:Bzicevib}
Upon adsorption of benzene on our proton-ordered ice surface model, the high symmetry of benzene is perturbed and most vibrational modes are now able to contribute to the infrared spectrum (and thus possibly indicating a $\mathcal{C}_1$ symmetry instead). Yet, in order to examine the correspondence between gas-phase and adsorbed modes, we analyse the vibrational modes in the $\mathcal{D}_{6h}$ point group when possible. 

\begin{figure}
\centering
\resizebox{0.75\columnwidth}{!}{
 \includegraphics{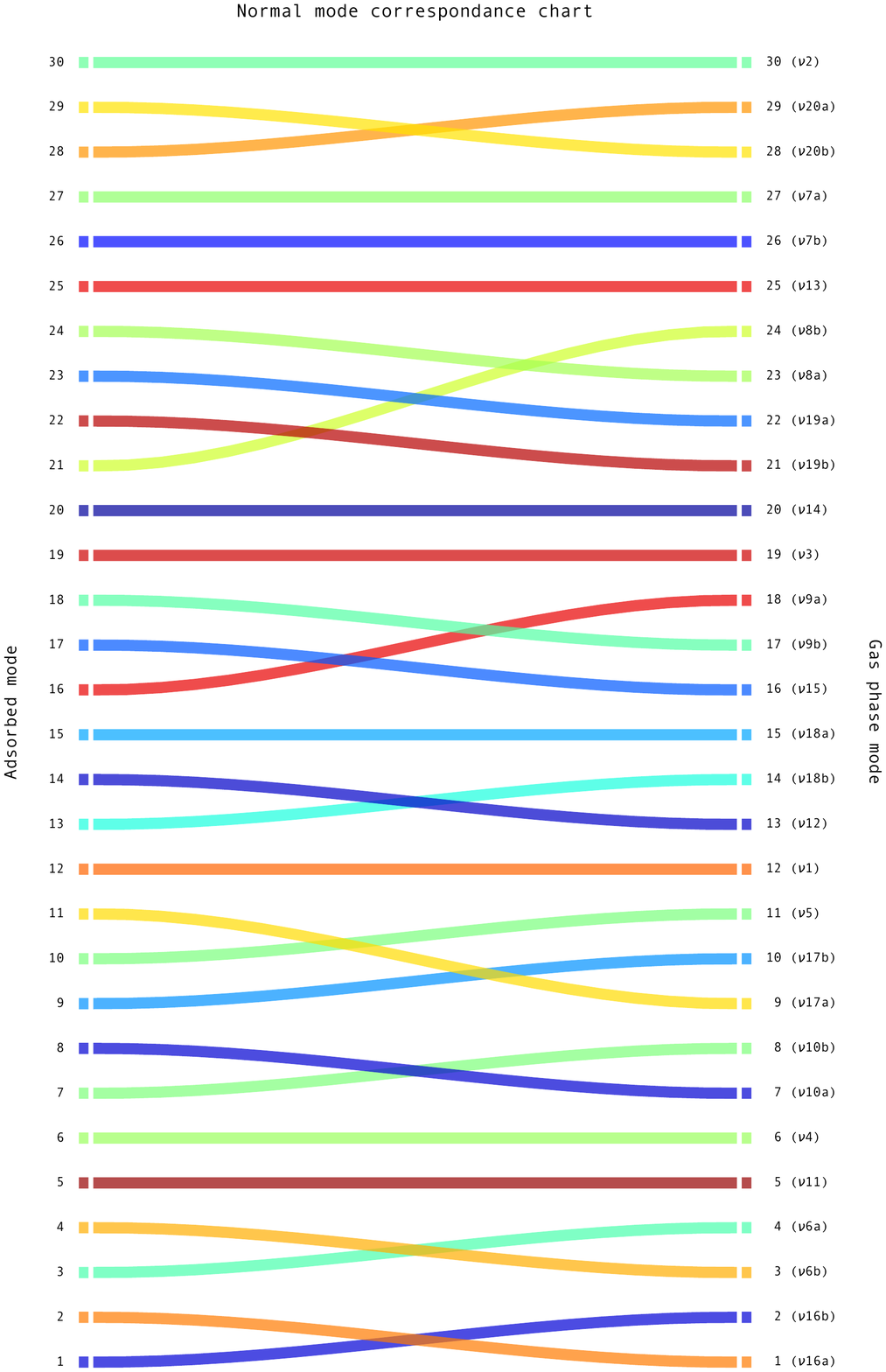}
}
\caption{The normal mode correspondence between benzene adsorbed onto ice surface (left) and gaseous benzene (right). The band labelling follow the convention adopted for gas-phase benzene in $\mathcal{D}_{6h}$ symmetry.}
\label{fig:benz_corr}
\end{figure}

To facilitate the cross-assignment between vibrational modes of $\mathcal{D}_{6h}$ gas-phase benzene and benzene adsorbed on the ice surface, we have analysed the overlap between each set of normal modes and generated the correspondence graph shown in Fig.~\ref{fig:benz_corr}. This figure shows the normal modes in gaseous benzene (right) mapped to the corresponding normal modes for adsorbed benzene onto an ordered ice surface (left). We see that while 10 modes retain their character upon adsorption (modes 5, 6, 12, 15, 19, 20, 25, 26, 27 and 30), most of the other modes are either swapped within their symmetry band (modes 1\&2, 3\&4, 7\&8, 13\&14 and 28\&29) or undergo a noticeable re-organisation. This is an indication of the influence of the ice surface on the vibrational structure of benzene, and is also important when comparing frequency shifts at a vibrational mode level (see below). 

Our result for the observed infrared bands are shown in Table \ref{tbl:bzicefreq}. For each band, we report results obtained using a harmonic model (PBE-D3 surface), our anharmonic model (using two potential energy surfaces) and the latest experimental results (a complete list of all frequencies is also shown in Table \ref{tab:booktabs-ice}). 

\begin{table}
 \centering
 \caption{Benzene on ice IR vibrational frequencies in \icm. All degenerate computed bands are intensity averaged (see Sec.~\ref{sec:intensity_average} for details). Weak bands observed by \citet{hagen_1983} are indicated in brackets. $\star$=($\nu_{20}$ in original reference), $\dag$=($\nu_{18}$ in original reference), $\ddag$ corresponds to a band that we have re-attributed based on \citet{hagen_1983}'s observations. RMSD (IR) only includes the modes that are originally infrared active for a ${\mathcal D}_{6{\rm h}}$ symmetry (bold modes), while RMSD (all) includes all bands.}
\begin{tabular}{cccccc}
\hline
\hline
& Harmonic & \multicolumn{2}{c}{Anharmonic} & & \\
\cline{3-4}
Mode [${\mathcal D}_{6{\rm h}}$ label] & PBE-D3 & PBE-D3 & CCSD(T)+PBE-D3 & \citet{hagen_1983} & \citet{dawes_2018} \\
\hline
{\bf 11} {[}A$_{\rm 2u}${]} & 619   & 657      & 677      & 681     & ---     \\
10 {[}E$_{\rm 1g}${]}    & 737   & 747      & 786      & 860     & 858     \\
17 {[}E$_{\rm 2u}${]}    & 848   & 798      & 843      & (977)    & 989     \\
12 {[}B$_{\rm 1u}${]}    & 991   & 975      & 1005     & (1012)   & 1011    \\
{\bf 18} {[}E$_{\rm 1u}${]} & 994   & 978      & 1002     & 1035$\star$    & 1035    \\
15 {[}B$_{\rm 2u}${]}    & 1133   & 1123      & 1144     & 1148    & 1148    \\
9 {[}E$_{\rm 1g}${]}    & 1106   & 1121      & 1149     & 1176    & 1177    \\
14 {[}B$_{\rm 2u}${]}    & 1353   & 1329      & 1328     & (1305)   & 1312    \\
{\bf 19} {[}E$_{\rm 1u}${]} & 1439   & 1413      & 1457     & 1478    & 1478    \\
8 {[}E$_{\rm 1g}${]}    & 1483   & 1419      & 1472     & (1585)   & 1586    \\
{\bf 20} {[}E$_{\rm 1u}${]} & 3084   & 2922      & 3020     & 3034$\dag$    & 3034    \\
7 {[}E$_{\rm 1g}${]}    & 3069   & 2943      & 3023     & (3050)   & ---    \\
2 {[}A$_{\rm 1g}${]}    & 3096   & 2925      & 3031     & 3069    & 3070$\ddag$    \\ \hline
RMSD (IR)       & 49    & 72       & 21      & ---     &       \\ 
RMSD (all)       & 68    & 100       & 56      & ---     &       \\ \hline
\hline
\end{tabular}
\label{tbl:bzicefreq}
\end{table}

\begin{figure}
\resizebox{\columnwidth}{!}{
 \includegraphics{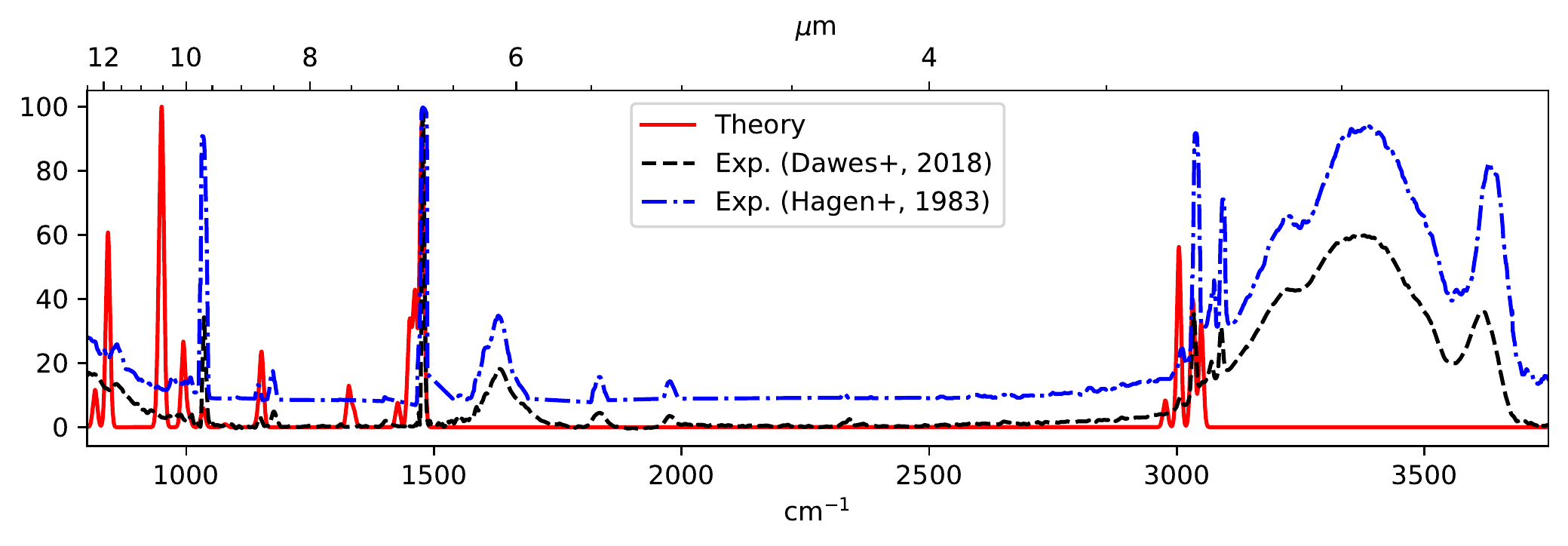}
}
  \caption{Comparison between simulated the IR spectrum for benzene on ferroelectric proton-ordered hexagonal crystalline water ice (XIh) surface model of this work (plain red trace); the experimental data from \citet{dawes_2018} (dashed black trace); and a digitised version of \citet{hagen_1983} (dot-dashed blue trace). See Sec. \ref{sec:Bzicevib} for computational details. All spectra are normalised so that their largest intensity peak corresponds to 100\%. The digitised trace was obtained from \citet{hagen_1983} using WebPlotDigitiser on their Figure 4 and, to compensate for artifacts in the digitisation, shifted by $+23$ \icm\ to match the reported values for benzene on ice in their Table V. Our shifted version of the digitised spectrum from \citet{hagen_1983} is available as a CSV file in supplementary material.}
\label{fig:ice-exp-comparison}
\end{figure}

At first glance, the two sets of experimental frequencies \citep{hagen_1983,dawes_2018} seem in very good agreement with each other. There are some discrepancies however, when we consider the weaker lines observed by \citet{hagen_1983} (indicated in brackets in Table~\ref{tbl:bzicefreq}) which are shifted by up to 12~\icm\ ($\nu_{17}$) compared to the measurements of \citet{dawes_2018}. On closer inspection of the original papers, we noticed a large deviation for $\nu_2$, which \citet{hagen_1983} attribute to their peak at 3069~\icm, while \citet{dawes_2018} attribute this band to a peak at 3089~\icm\ in their spectrum. This assignment would lead to a large disagreement of 20~\icm\ for this strong band --- unlike all other strong bands observed. It is worth noting that \citet{hagen_1983} do record a peak at 3090~\icm\ (see also the spectra shown in Fig.\ref{fig:ice-exp-comparison}) but attribute this to a combination band ($\nu_8+\nu_{19}$) of significant intensity. Given the overall excellent agreement between the two studies for the strong intensity modes, we are inclined to think that \citet{dawes_2018} band $\nu_7$ is more likely to correspond to band $\nu_2$ and that $\nu_7$ (a band described as weak by \citet{hagen_1983}) is not present in their spectra. Another issue stems from the ambiguous assignment of $\nu_{2}$ and $\nu_{13}$ in \citet{hagen_1983} (see also Table \ref{tab:booktabs-ice}). In their paper, the peak at 3069~\icm\ discussed above is attributed to either band $\nu_{13}$ or $\nu_{2}$ (Table V in \citet{hagen_1983} listing it as ``13 or 2''). This assignment probably originates from a comparison with the comprehensive work of \citet{szczepaniak_1972} on benzene in a solid HCl matrix where they list a $\nu_{13}$ band at 3063~\icm\ and indicate the $\nu_2$ band as a possible shoulder around 3051~\icm. Based on the later work of \citet{dawes_2018} and our own calculations, it seems more likely that the peak at 3069~\icm\ in \citet{hagen_1983} corresponds to $\nu_2$ as it is a stronger intensity carrier than $\nu_{13}$.

Analysing the computational results from Table \ref{tbl:bzicefreq}, we see that the strong bands are qualitatively well described by all models. The harmonic approximation underestimates slightly the bands below about 1200~\icm\ ($\nu_{11}$, $\nu_{10}$, $\nu_{17}$, $\nu_{12}$, $\nu_{18}$, $\nu_{15}$ and $\nu_9$) and overestimates the C--H stretching bands ($\nu_{20}$, $\nu_7$ and $\nu_2$). This indicates that a global scaling of the harmonic frequencies (often used to account for anharmonicity in an empirical fashion, see also \citet{Kesharwani_2014}) is unlikely to lead to a satisfactory agreement here. The largest deviations from the experimental data of \citet{hagen_1983} are seen for bands $\nu_{10}$ (123~\icm), $\nu_{17}$ (129~\icm), and $\nu_8$ (102~\icm), with slightly larger deviations compared to \citet{dawes_2018}. Overall, the harmonic approximation has a root-mean-square deviation (RMSD) of 49~\icm\ for the original gas-phase IR active modes ($\nu_{11}$, $\nu_{18}$, $\nu_{19}$ and $\nu_{20}$, shown in bold in Table ~\ref{tbl:bzicefreq}) and 68~\icm\ for all observed bands, compared to the data of \citet{hagen_1983}.

The anharmonic approach underestimates all bands apart from $\nu_{14}$. This is more pronounced for the PBE-D3 surface, with the largest deviation for $\nu_{10}$ (113~\icm), $\nu_{17}$ (179~\icm), $\nu_{8}$ (166~\icm), $\nu_{20}$ (112~\icm), $\nu_7$ (107~\icm) and $\nu_2$ (144~\icm). Note that all C--H stretches are underestimated by over 100~\icm\ for this surface, similarly to the PBE-D3 results for gas-phase benzene. We see the RMSD increase for this PES to 72~\icm\ for the original IR active bands and 100~\icm\ for all observed bands. 

Our hybrid PES improves the situation significantly by providing a ``steeper'' potential energy surface that closes some of the gaps between theory and experiment. The overall agreement is now much improved, with an RMSD of 21~\icm\ for the original IR active bands and 56~\icm\ for all observed bands, outperforming both the PBE-D3 surface and the harmonic approximation. The largest deviations are seen for bands $\nu_{17}$ (134~\icm) and $\nu_{8}$ (113~\icm). All C--H stretching bands are improved by an order of magnitude compared to the PBE-D3 PES data. 

We compare our best first-principle model spectrum with that of \citet{hagen_1983} and \citet{dawes_2018} in Figure \ref{fig:ice-exp-comparison}. The simulated spectrum is obtained from the hybrid vibrational frequencies and PBE-D3 intensities shown in Table \ref{tab:booktabs-ice} convolved with a Gaussian lineshape with 5~\icm\ half-width at half-maximum (HWHM), which mimics the linewidth of \citet{hagen_1983}. We see that our predicted spectrum agrees qualitatively very well with experiment, displaying a structure for the C--H stretching bands similar to that observed experimentally. The strong band just below 1500~\icm\ ($\nu_{8}$ and $\nu_{19}$) is well reproduced along with other low frequency peaks. The overall spectrum is red shifted in parts for reasons discussed later, but overall it provides a suitable guide to explore the nature of the bands observed for benzene on an ice surface. 

\begin{figure}
\resizebox{\columnwidth}{!}{
 \includegraphics{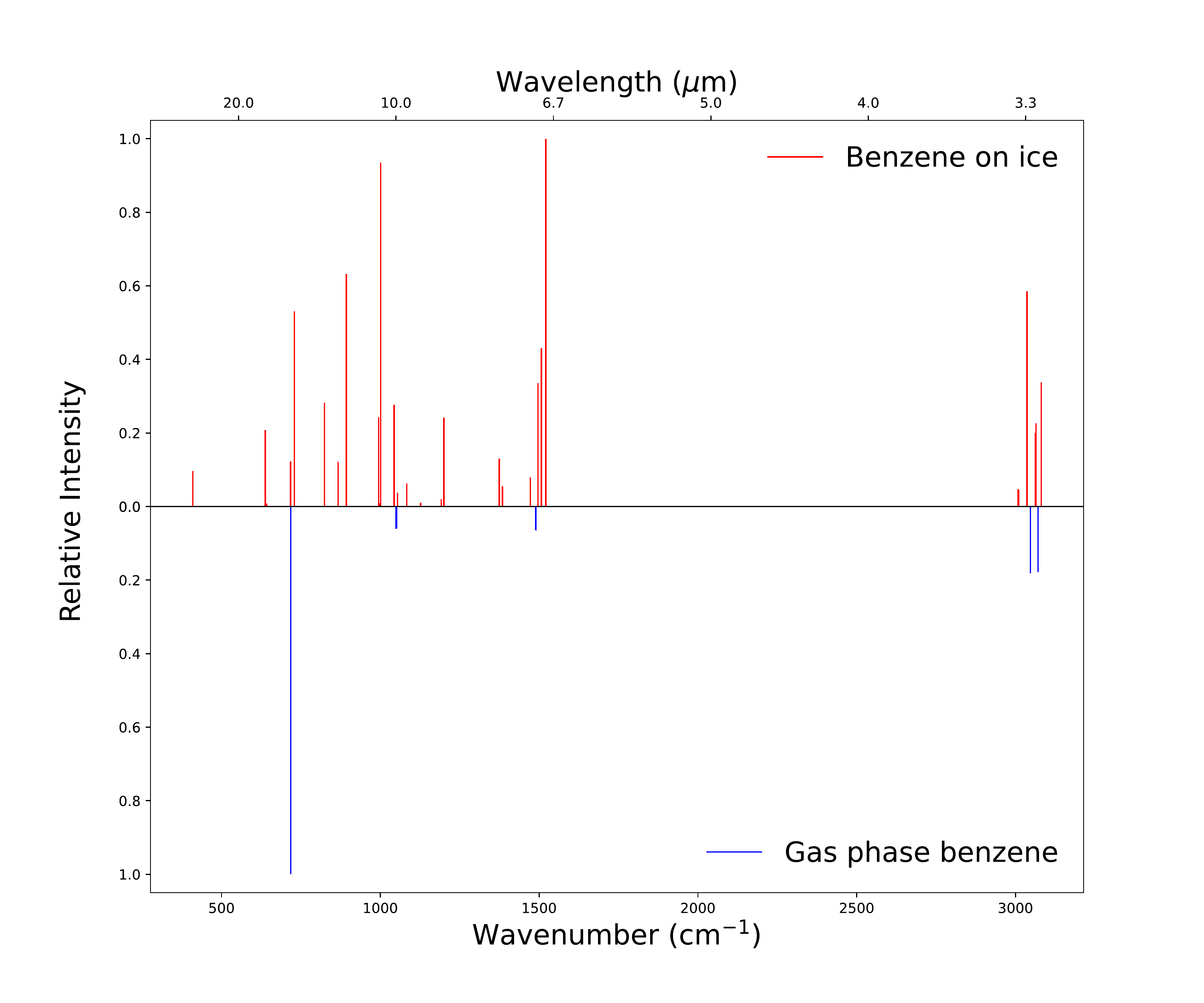}
}
\caption{Relative intensities and frequencies of the IR visible transitions for benzene adsorbed onto ice surface (top, red) and gaseous benzene (bottom, blue). The stick spectra are generated using data from both Table~\ref{tab:booktabs-gas} and Table~\ref{tab:booktabs-ice}. It is clear that the number, position and relative intensities of the IR peaks change drastically when the benzene is adsorbed onto a model ice surface.}
\label{fig:ice-gas-comparison}
\end{figure}

Figure \ref{fig:ice-gas-comparison} shows the intensity and position of all computed IR visible peaks for benzene on ice (top) and gaseous benzene (bottom), using data from both Table~\ref{tab:booktabs-gas} and Table~\ref{tab:booktabs-ice}. This qualitative figure emphasises how the number, position, and relative intensities of the peaks change when the benzene is adsorbed on a water ice surface compared to when it is its gaseous form. Our anharmonic model captures correctly the overall shape and spectral changes going from gas-phase benzene, with only 4 active infrared bands, to adsorbed benzene on ice which displays approximately 13 bands in the infrared. Moreover, we see the overall intensity pattern shifts from a dominating $\nu_{11}$ band in gas-phase benzene (used as identification fingerprint in CRL618 by \citep{cernicharo_2001}, for example) to several high-intensity carriers for adsorbed benzene ($\nu_{8}$ and $\nu_{19}$, for example). 

The near-global underestimation of the vibrational bands for the anharmonic calculations are likely an indication of the strength of the interaction between benzene and the ice surface. Indeed, the stronger the interaction, the more affected the vibrational modes are and, consequently, the more red shifted (lower frequency) they will be. As the benzene--ice interaction is computed at the same level of theory for all models (using PBE-D3), we expect the trends observed to have a common source. The PBE-D3 approach has been shown to overestimate the binding energy of molecules on surfaces (\citet{reckien_2014} for benzene adsorption on metal surfaces, for example) which could explain the excessive red shift seen in our theoretical values. 
\begin{table}
 \centering
 \caption{Vibrational frequency shift, $\Delta \nu$, for benzene adsorbed on ice compared to gas-phase in \icm. We define $\Delta\nu_i=\nu_i^{\textrm{gas}}-\nu_i^{\textrm{ice}}$. All degenerate computed bands have been intensity averaged (see Sec.~\ref{sec:intensity_average} for details). Modes that are originally infrared active for a ${\mathcal D}_{6{\rm h}}$ symmetry are shown in bold. Experimental shifts are computed using the experimental values reported for the gas phase in Table \ref{tab:booktabs-gas} and the theoretical values are computed similarly using the gas-phase values obtained for the corresponding level of theory.}
\begin{tabular}{cccccc}
\hline
\hline
& Harmonic & \multicolumn{2}{c}{Anharmonic} & & \\
\cline{3-4}
Mode [${\mathcal D}_{6{\rm h}}$ label] & PBE-D3 & PBE-D3 & CCSD(T)+PBE-D3 & \citet{hagen_1983} & \citet{dawes_2018} \\
\hline
{\bf 11} {[}A$_{\rm 2u}${]} & $+44   $ & $+37      $ & $+41     $ & $-7 $ & --- \\
10 {[}E$_{\rm 1g}${]}    & $+97   $ & $+99      $ & $+93     $ & $-13$ & $-11$ \\
17 {[}E$_{\rm 2u}${]}    & $+100  $ & $+155     $ & $+151     $ & $-10$ & $-22$ \\
12 {[}B$_{\rm 1u}${]}    & $+4   $ & $+9      $ & $+11     $ & $-2 $ & $-1$ \\
{\bf 18} {[}E$_{\rm 1u}${]} & $+40   $ & $+48      $ & $+49     $ & $+3 $ & $+3$ \\
15 {[}B$_{\rm 2u}${]}    & $+10   $ & $+19      $ & $+24     $ & $0 $ & $0$  \\
9 {[}E$_{\rm 1g}${]}    & $+56   $ & $+33      $ & $+35     $ & $+2 $ & $+1$ \\
14 {[}B$_{\rm 2u}${]}    & $-6   $ & $-3      $ & $-13     $ & $+4 $ & $-3$ \\
{\bf 19} {[}E$_{\rm 1u}${]} & $+26   $ & $+30      $ & $+33     $ & $+6 $ & $+6$ \\
8 {[}E$_{\rm 1g}${]}    & $+108  $ & $+147     $ & $+148     $ & $+16$ & $+15$ \\
{\bf 20} {[}E$_{\rm 1u}${]} & $+21   $ & $+27      $ & $+39     $ & $+14$ & $+14$ \\
7 {[}E$_{\rm 1g}${]}    & $+22   $ & $-7      $ & $+18     $ & $+7 $ & --- \\
2 {[}A$_{\rm 1g}${]}    & $+19   $ & $+57      $ & $+50     $ & $+5 $ & $+4$ \\
\hline
Average, $\langle\Delta\nu\rangle$    & $+42   $ & $+50      $ & $+52     $ & $+2 $ & $+1$ \\
RMSD     & $+55   $ & $+71      $ & $+71     $ & $+8 $ & $+10$\\
\hline
\hline
\end{tabular}
\label{tbl:bziceshift}
\end{table}

In order to gain further insights into the influence of surface adsorption on the vibrational spectrum of benzene, we computed the frequency shift, $\Delta \nu$ resulting from adsorption. We define this frequency shift as:
\begin{eqnarray}
\Delta\nu_i=\nu_i^{\textrm{gas}}-\nu_i^{\textrm{ice}}
\end{eqnarray}
where $\nu_i^{\textrm{gas}}$ is the frequency of band $\nu_i$ for benzene in the gas phase and $\nu_i^{\textrm{ice}}$ that for adsorbed benzene. The experimental shifts are computed using the values of \citet{hagen_1983} and \citet{dawes_2018}, along with the corresponding experimental frequencies for gas-phase benzene (see Table \ref{tab:booktabs-gas}). Our results are shown in Table \ref{tbl:bziceshift}. We observe a slight blue shift (adsorbed benzene bands occuring at higher frequencies) for out-of-plane bands ($\nu_{11}$, $\nu_{10}$ and $\nu_{17}$), while most other bands are either weakly affected ($\nu_{12}$, $\nu_{18}$, $\nu_{15}$, $\nu_{9}$ and $\nu_{14}$) or red shifted by up to $+16$~\icm\ ($\nu_{19}$, $\nu_{8}$, $\nu_{20}$, $\nu_{7}$ and $\nu_{2}$). Blue shifting is consistent with a stiffening of out-of-plane modes for an adsorbed molecule that vibrates against a fixed (slightly repulsive) surface (e.g. normal mode corresponding to $\nu_{11}$ shown in Figure~\ref{fig:bzmodes} B, and also \citet{zamirri_2018} for a similar observation for CO on water ice). Red shifting, on the other hand, is indicative of an attractive interaction between benzene and the water molecules, typically through hydrogen bonding (see for example \citet{MUKHOPADHYAY_2015} for similar observations for the water dimer). 

Our theoretical shift values are computed similarly using the gas-phase values obtained for the each corresponding level of theory. The vibrational shift provide an information that is typically more reliable than absolute frequency values. Indeed, taking the frequency difference can cancel systematic method issues, as long as both gas-phase and adsorbed frequencies are computed using the same method (see also \citet{michoulier_2020} where a similar approach is used to explore water ice OH-stretch shifts). Here, we see that all three theoretical approaches show a much larger shift than the two experimental studies. 

Interestingly, we notice that all three models follow a similar pattern with mostly positive shifts (i.e.\ frequencies are lower (red shifted) for adsorbed benzene compared to the gas-phase) apart from $\Delta\nu_{14}$ which is slightly negative. The anharmonic PBE-D3 results also display a negative shift for $\nu_7$, likely due to the issues mentioned earlier regarding the description of stretching modes in pure DFT. One particular qualitative feature is that the low-frequency out-of-plane blue shifts observed experimentaly are not reproduced by any of the approaches. The remaining modes, however, mirror roughly the experimental trend albeit with a much larger magnitude.

To investigate the overall magnitude of the deviations, we explore the average shift, $\langle\Delta\nu\rangle$, and root-mean-square deviation (RMSD) for all reported vibrational shifts in Table \ref{tbl:bziceshift}. The harmonic model has an average shift of $+42$~\icm\ and corresponding RMSD of $+55$~\icm. Similarly, the anharmonic model provides an average of $+50$~\icm\ with the PBE-D3 surface ($+71$~\icm\ RMSD) and a $+52$~\icm\ for the hybrid surface ($+71$~\icm\ RMSD). Those average shifts are to be compared with much lower $\langle\Delta\nu\rangle$ for the experimental data ($+2$~\icm\ for \citet{hagen_1983} and $+1$ for \citet{dawes_2018}) and RMSD values ($+8$~\icm\ and $+10$~\icm, respectively). Here again, we see that either measure (average or RMSD) of the shift magnitude is much larger for the models than what is computed for the experimental data. This is another indication that the interaction of benzene with the ice surface is likely overestimated by the PBE-D3 approach. Another possible explanation is that our model represents the most binding version of an ice surface, while the experiments were performed using low-temperature deposited disordered ice layers (ASW) that is likely less strongly interacting with benzene.

\section{Conclusion}
\label{sec:conc}
In this paper we have presented a hybrid approach to calculating the vibrational signatures for a system of benzene adsorbed onto a water-ice surface by using a combination of CCST(T) and PBE-D3 in an ONIOM-like procedure. We compare the results of our hybrid method with the latest experimentally recorded spectra and calculations performed using either harmonic or anharmonic methods using a PBE-D3-only PES. When used for gaseous benzene calculations, the hybrid approach achieved an RMSD for IR active modes of 24~\icm, compared to 55~\icm\ and 31\icm\ for the anharmonic and harmonic PBE-D3-only approaches, respectively. 

We used a proton-ordered ferroelectric XIh water-ice surface model consisting of 288 water molecules arranged in four double-layers of H$_{2}$O molecules, and our hybrid calculated spectrum for benzene on ice agrees well with the experimentally measured spectra of \citet{hagen_1983} and \citet{dawes_2018} on ASW, resulting in a RMSD for IR active modes of 21~\icm, compared to 72~\icm\ and 49\icm\ for the anharmonic PBE-D3 approach and the harmonic PBE-D3 approaches, respectively. Our hybrid approach remains relatively ``low-cost" compared to a full correlation treatment and thus is suitable for the investigation of larger aromatic species on water ice surfaces. 

We have also shown the importance of including a water ice surface into spectroscopic calculations which lead to differences in vibrational modes, vibrational frequencies, intensities and number of transitions seen in the IR spectrum. The overall intensity pattern of the spectrum changed from a being dominating by the $\nu_{11}$ band in gas-phase benzene to showing several high-intensity carriers for adsorbed benzene. To facilitate the cross-assignment between vibrational modes of gaseous benzene versus benzene on an ice surface, we have analysed the overlap between each set of normal modes and generated a correspondence graph.

We calculate the frequency shifts, $\Delta\nu$, of adsorbed benzene relative to its gas phase for both experimental data and our theoretical models. We show that, while there is a reasonable common agreement between harmonic and anharmonic approaches, the underlying description of the benzene--water interactions at the PBE-D3 level is likely too strong and leads to an excessive red shift for all methods.

Finally, both sets of IR measurements were performed on amorphous structured water (ASW) ice but we find here that there is a surprising agreement with our computed highly-ordered crystalline ice data. This implies either that the simple ordered model that we developed describes realistically the local environment experienced by benzene on ASW (i.e.\ on a local scale, the binding mode of benzene in ASW resembles that of ice XIh). We are currently exploring this avenue further and will report our findings in a future publication.

\section{Acknowledgements}
We thank Dr Anita Dawes (The Open University, UK) for generously providing experimental data used in this work. 
We acknowledge the Viper High Performance Computing facility of the University of Hull and its support team.

\bibliographystyle{mnras.bst}
\bibliography{paper}

\begin{thebibliography}{}
\makeatletter
\relax
\def\mn@urlcharsother{\let\do\@makeother \do\$\do\&\do\#\do\^\do\_\do\%\do\~}
\def\mn@doi{\begingroup\mn@urlcharsother \@ifnextchar [ {\mn@doi@}
  {\mn@doi@[]}}
\def\mn@doi@[#1]#2{\def\@tempa{#1}\ifx\@tempa\@empty \href
  {http://dx.doi.org/#2} {doi:#2}\else \href {http://dx.doi.org/#2} {#1}\fi
  \endgroup}
\def\mn@eprint#1#2{\mn@eprint@#1:#2::\@nil}
\def\mn@eprint@arXiv#1{\href {http://arxiv.org/abs/#1} {{\tt arXiv:#1}}}
\def\mn@eprint@dblp#1{\href {http://dblp.uni-trier.de/rec/bibtex/#1.xml}
  {dblp:#1}}
\def\mn@eprint@#1:#2:#3:#4\@nil{\def\@tempa {#1}\def\@tempb {#2}\def\@tempc
  {#3}\ifx \@tempc \@empty \let \@tempc \@tempb \let \@tempb \@tempa \fi \ifx
  \@tempb \@empty \def\@tempb {arXiv}\fi \@ifundefined
  {mn@eprint@\@tempb}{\@tempb:\@tempc}{\expandafter \expandafter \csname
  mn@eprint@\@tempb\endcsname \expandafter{\@tempc}}}

\bibitem[\protect\citeauthoryear{Acke \& van~den Ancker}{Acke \& van~den
  Ancker}{2004}]{acke_2004}
Acke B.,  van~den Ancker M.~E.,  2004, \mn@doi [Astron. Astrophys.]
  {10.1051/0004-6361:20040400}, 426, 151

\bibitem[\protect\citeauthoryear{Akima}{Akima}{1996}]{akima_1996}
Akima H.,  1996, \mn@doi [{ACM} Trans. Math. Softw.] {10.1145/232826.232856},
  22, 362

\bibitem[\protect\citeauthoryear{Arakawa, Kagi  \& Fukazawa}{Arakawa
  et~al.}{2009}]{arakawa_2009}
Arakawa M.,  Kagi H.,   Fukazawa H.,  2009, \mn@doi [Astrophys. J., Suppl.
  Ser.] {10.1088/0067-0049/184/2/361}, 184, 361

\bibitem[\protect\citeauthoryear{Augspurger, Dykstra  \& Zwier}{Augspurger
  et~al.}{1992}]{augspurger_1992}
Augspurger J.~D.,  Dykstra C.~E.,   Zwier T.~S.,  1992, \mn@doi [J. Phys.
  Chem.] {10.1021/j100197a023}, 96, 7252

\bibitem[\protect\citeauthoryear{Balint-Kurti, Ward  \&
  Clay~Marston}{Balint-Kurti et~al.}{1991}]{balint-kurti_1991}
Balint-Kurti G.~G.,  Ward C.~L.,   Clay~Marston C.,  1991, \mn@doi [Comput.
  Phys. Commun.] {10.1016/0010-4655(91)90023-E}, 67, 285

\bibitem[\protect\citeauthoryear{Benoit}{Benoit}{2015}]{benoit_2015}
Benoit D.~M.,  2015, \mn@doi [J. Phys. Chem. A] {10.1021/acs.jpca.5b08543},
  119, 11583

\bibitem[\protect\citeauthoryear{Benoit, Chavagnac  \& Clary}{Benoit
  et~al.}{1998}]{benoit_1998}
Benoit D.~M.,  Chavagnac A.~X.,   Clary D.~C.,  1998, \mn@doi [Chem. Phys.
  Lett.] {10.1016/S0009-2614(97)01396-1}, 283, 269

\bibitem[\protect\citeauthoryear{Benoit, Madebene, Ulusoy, Mancera, Scribano
  \& Chulkov}{Benoit et~al.}{2011}]{benoit_2011}
Benoit D.~M.,  Madebene B.,  Ulusoy I.,  Mancera L.,  Scribano Y.,   Chulkov
  S.,  2011, \mn@doi [Beilstein J. Nanotechnol.] {10.3762/bjnano.2.48}, 2, 427

\bibitem[\protect\citeauthoryear{Brodersen \& Langseth}{Brodersen \&
  Langseth}{1956}]{brodersen_1956}
Brodersen S.,  Langseth A.,  1956, The Infrared Spectra of Benzene,
  Sym-benzene-d3, and Benzene-d6.
E. Munksgaard

\bibitem[\protect\citeauthoryear{Brodersen \& Langseth}{Brodersen \&
  Langseth}{1959}]{brodersen_1959}
Brodersen S.,  Langseth A.,  1959, Kgl. Danske Videnskab. Selskab, Mat.-fys.
  Skrifter, 1

\bibitem[\protect\citeauthoryear{Cernicharo, Heras, Tielens, Pardo, Herpin,
  Gu\'{e}lin  \& Waters}{Cernicharo et~al.}{2001}]{cernicharo_2001}
Cernicharo J.,  Heras A.~M.,  Tielens A. G. G.~M.,  Pardo J.~R.,  Herpin F.,
  Gu\'{e}lin M.,   Waters L. B. F.~M.,  2001, \mn@doi [Astrophys. J.]
  {10.1086/318871}, 546, L123

\bibitem[\protect\citeauthoryear{Chergui}{Chergui}{1996}]{chergui_1996}
Chergui M.,  1996, Femtochemistry: Ultrafast chemical and physical processes in
  molecular systems.
{World} {Scientific}

\bibitem[\protect\citeauthoryear{Christiansen, Stanton  \& Gauss}{Christiansen
  et~al.}{1998}]{christiansen_1998}
Christiansen O.,  Stanton J.~F.,   Gauss J.,  1998, \mn@doi [J. Chem. Phys.]
  {10.1063/1.475801}, 108, 3987

\bibitem[\protect\citeauthoryear{Chulkov \& Benoit}{Chulkov \&
  Benoit}{2013}]{chulkov_2013}
Chulkov S.~K.,  Benoit D.~M.,  2013, \mn@doi [J. Chem. Phys.]
  {10.1063/1.4829461}, 139, 214704

\bibitem[\protect\citeauthoryear{Clark \& Benoit}{Clark \&
  Benoit}{2019}]{clark_2019}
Clark V.~H.,  Benoit D.~M.,  2019, \mn@doi [International Astronomical Union.
  Proceedings of the International Astronomical Union]
  {10.1017/S174392131900944X}, 15, 468

\bibitem[\protect\citeauthoryear{Dapprich, Kom{\`a}romi, Byun, Morokuma  \&
  Frisch}{Dapprich et~al.}{1999}]{dapprich_1999}
Dapprich S.,  Kom{\`a}romi I.,  Byun K.,  Morokuma K.,   Frisch M.~J.,  1999,
  \mn@doi [J. Mol. Struct.] {10.1016/S0166-1280(98)00475-8}, 461--462, 1

\bibitem[\protect\citeauthoryear{Dawes, Pascual, Mason, G\"{a}rtner, Hoffmann
  \& Jones}{Dawes et~al.}{2018}]{dawes_2018}
Dawes A.,  Pascual N.,  Mason N.~J.,  G\"{a}rtner S.,  Hoffmann S.~V.,   Jones
  N.~C.,  2018, \mn@doi [Phys. Chem. Chem. Phys.] {10.1039/C8CP01228H}, 20,
  15273

\bibitem[\protect\citeauthoryear{Delsemme}{Delsemme}{1975}]{delsemme_1975}
Delsemme A.,  1975, \mn@doi [Icarus]
  {https://doi.org/10.1016/0019-1035(75)90163-3}, 24, 95

\bibitem[\protect\citeauthoryear{Dunning}{Dunning}{1971}]{dunning_1971}
Dunning T.~H.,  1971, \mn@doi [J. Chem. Phys.] {10.1063/1.1676139}, 55, 716

\bibitem[\protect\citeauthoryear{Dunning}{Dunning}{1989}]{dunning_1989}
Dunning T.~H.,  1989, \mn@doi [J. Chem. Phys.] {10.1063/1.456153}, 90, 1007

\bibitem[\protect\citeauthoryear{Engdahl \& Nelander}{Engdahl \&
  Nelander}{1985}]{engdahl_1985}
Engdahl A.,  Nelander B.,  1985, \mn@doi [J. Phys. Chem.]
  {10.1021/j100259a031}, 89, 2860

\bibitem[\protect\citeauthoryear{Erlekam, Frankowski, Meijer  \& von
  Helden}{Erlekam et~al.}{2006}]{erlekam_2006}
Erlekam U.,  Frankowski M.,  Meijer G.,   von Helden G.,  2006, \mn@doi [J.
  Chem. Phys.] {10.1063/1.2198828}, 124, 171101

\bibitem[\protect\citeauthoryear{Faraday}{Faraday}{1825}]{faraday_1825}
Faraday M.,  1825, \mn@doi [Philos. Trans. Roy. Soc.] {10.1098/rstl.1825.0022},
  1, 440

\bibitem[\protect\citeauthoryear{Fukazawa, Hoshikawa, Ishii, Chakoumakos  \&
  Fernandez-Baca}{Fukazawa et~al.}{2006}]{fukazawa_2006}
Fukazawa H.,  Hoshikawa A.,  Ishii Y.,  Chakoumakos B.~C.,   Fernandez-Baca
  J.~A.,  2006, \mn@doi [Astrophys. J.] {10.1086/510017}, 652, L57

\bibitem[\protect\citeauthoryear{Geers et~al.,}{Geers
  et~al.}{2006}]{geers_2006}
Geers V.,  et~al., 2006, \mn@doi [Astron. Astrophys.]
  {10.1051/0004-6361:20064830}, 459, 545

\bibitem[\protect\citeauthoryear{Goedecker, Teter  \& Hutter}{Goedecker
  et~al.}{1996}]{goedecker_1996}
Goedecker S.,  Teter M.,   Hutter J.,  1996, \mn@doi [Phys. Rev. B]
  {10.1103/PhysRevB.54.1703}, 54, 1703

\bibitem[\protect\citeauthoryear{Goodman, Berman  \& Ozkabak}{Goodman
  et~al.}{1989}]{goodman-1989}
Goodman L.,  Berman J.~M.,   Ozkabak A.~G.,  1989, \mn@doi [J. Chem. Phys.]
  {10.1063/1.455951}, 90, 2544

\bibitem[\protect\citeauthoryear{Gotch \& Zwier}{Gotch \&
  Zwier}{1992}]{gotch_1992}
Gotch A.~J.,  Zwier T.~S.,  1992, \mn@doi [J. Chem. Phys.] {10.1063/1.461940},
  96, 3388

\bibitem[\protect\citeauthoryear{{Hagen}, {Tielens}  \& {Greenberg}}{{Hagen}
  et~al.}{1983}]{hagen_1983}
{Hagen} W.,  {Tielens} A.~G.~G.~M.,   {Greenberg} J.~M.,  1983, Astron.
  Astrophys. Suppl., 51, 389

\bibitem[\protect\citeauthoryear{Hirsch \& Ojam\"{a}e}{Hirsch \&
  Ojam\"{a}e}{2004}]{hirsch_2004}
Hirsch T.~K.,  Ojam\"{a}e L.,  2004, \mn@doi [J. Phys. Chem. B]
  {10.1021/jp048434u}, 108, 15856

\bibitem[\protect\citeauthoryear{Hollenstein, Piccirillo, Quack  \&
  Snels}{Hollenstein et~al.}{1990}]{hollenstein_1990}
Hollenstein H.,  Piccirillo S.,  Quack M.,   Snels M.,  1990, \mn@doi [Mol.
  Phys.] {10.1080/00268979000102091}, 71, 759

\bibitem[\protect\citeauthoryear{Hollinger \& Welsh}{Hollinger \&
  Welsh}{1978}]{hollinger_1978}
Hollinger A.~B.,  Welsh H.~L.,  1978, \mn@doi [Can. J. Phys.]
  {10.1139/p78-202}, 56, 1513

\bibitem[\protect\citeauthoryear{Jensen \& Brodersen}{Jensen \&
  Brodersen}{1979}]{jensen_1979}
Jensen H.~B.,  Brodersen S.,  1979, \mn@doi [J. Raman Spectrosc.]
  {10.1002/jrs.1250080208}, 8, 103

\bibitem[\protect\citeauthoryear{Jones, Zhang, Kaiser, Jamal, Mebel, Cordiner
  \& Charnley}{Jones et~al.}{2011}]{jones_2011}
Jones B.~M.,  Zhang F.,  Kaiser R.~I.,  Jamal A.,  Mebel A.~M.,  Cordiner
  M.~A.,   Charnley S.~B.,  2011, \mn@doi [Proc. Nat. Acad. Sci.]
  {10.1073/pnas.1012468108}, 108, 452

\bibitem[\protect\citeauthoryear{Jung \& Gerber}{Jung \&
  Gerber}{1996}]{jung_1996}
Jung J.~O.,  Gerber R.~B.,  1996, \mn@doi [J. Chem. Phys.] {10.1063/1.472960},
  105, 10332

\bibitem[\protect\citeauthoryear{Jure\v{c}ka, \v{S}poner, \v{C}ern\'{y}  \&
  Hobza}{Jure\v{c}ka et~al.}{2006}]{jurecka_2006}
Jure\v{c}ka P.,  \v{S}poner J.,  \v{C}ern\'{y} J.,   Hobza P.,  2006, \mn@doi
  [Phys. Chem. Chem. Phys.] {10.1039/B600027D}, 8, 1985

\bibitem[\protect\citeauthoryear{Kaiser, Parker  \& Mebel}{Kaiser
  et~al.}{2015}]{kaiser_2015}
Kaiser R.~I.,  Parker D.~S.,   Mebel A.~M.,  2015, \mn@doi [Ann. Rev. Phys.
  Chem.] {10.1146/annurev-physchem-040214-121502}, 66, 43

\bibitem[\protect\citeauthoryear{Kayi, Kaiser  \& Head}{Kayi
  et~al.}{2011}]{kayi_2011}
Kayi H.,  Kaiser R.~I.,   Head J.~D.,  2011, \mn@doi [Phys. Chem. Chem. Phys.]
  {10.1039/c1cp20656g}, 13, 15774

\bibitem[\protect\citeauthoryear{Ke\c{c}eli, Hirata  \& Yagi}{Ke\c{c}eli
  et~al.}{2010}]{keceli_2010}
Ke\c{c}eli M.,  Hirata S.,   Yagi K.,  2010, \mn@doi [J. Chem. Phys.]
  {10.1063/1.3462238}, 133, 034110

\bibitem[\protect\citeauthoryear{Kendall, Dunning  \& Harrison}{Kendall
  et~al.}{1992}]{kendall_1992}
Kendall R.~A.,  Dunning T.~H.,   Harrison R.~J.,  1992, \mn@doi [J. Chem.
  Phys.] {10.1063/1.462569}, 96, 6796

\bibitem[\protect\citeauthoryear{Kesharwani, Brauer  \& Martin}{Kesharwani
  et~al.}{2015}]{Kesharwani_2014}
Kesharwani M.~K.,  Brauer B.,   Martin J. M.~L.,  2015, \mn@doi [J. Phys. Chem.
  A] {10.1021/jp508422u}, 119, 1701

\bibitem[\protect\citeauthoryear{Koskinen, Moses, West, Guerlet  \&
  Jouchoux}{Koskinen et~al.}{2016}]{Koskinen_2016}
Koskinen T.~T.,  Moses J.~I.,  West R.~A.,  Guerlet S.,   Jouchoux A.,  2016,
  \mn@doi [Geophys. Res. Lett.] {10.1002/2016GL070000}, 43, 7895

\bibitem[\protect\citeauthoryear{Krack}{Krack}{2005}]{krack_2005}
Krack M.,  2005, \mn@doi [Theor. Chem. Acc.] {10.1007/s00214-005-0655-y}, 114,
  145

\bibitem[\protect\citeauthoryear{Lippert, Hutter  \& Parrinello}{Lippert
  et~al.}{1997}]{lippert_1997}
Lippert G.,  Hutter J.,   Parrinello M.,  1997, \mn@doi [Mol. Phys.]
  {10.1080/00268979709482119}, 92, 477

\bibitem[\protect\citeauthoryear{Ma \& Ma}{Ma \& Ma}{2012}]{ma_2012}
Ma H.,  Ma Y.,  2012, \mn@doi [J. Chem. Phys.] {10.1063/1.4769124}, 137, 214504

\bibitem[\protect\citeauthoryear{Marston \& Balint-Kurti}{Marston \&
  Balint-Kurti}{1989}]{marston_1989}
Marston C.~C.,  Balint-Kurti G.~G.,  1989, \mn@doi [J. Chem. Phys.]
  {10.1063/1.456888}, 91, 3571

\bibitem[\protect\citeauthoryear{Matsuura et~al.,}{Matsuura
  et~al.}{2004}]{matsuura_2004}
Matsuura M.,  et~al., 2004, \mn@doi [Astrophys. J.] {10.1086/382064}, 604, 791

\bibitem[\protect\citeauthoryear{{McKinnon} \& {Hofmeister}}{{McKinnon} \&
  {Hofmeister}}{2005}]{McKinnon_2005}
{McKinnon} W.~B.,  {Hofmeister} A.~M.,  2005, Bull. Am. Astron. Soc., p.~732

\bibitem[\protect\citeauthoryear{Michoulier, Toubin, Simon, Mascetti, Aupetit
  \& Noble}{Michoulier et~al.}{2020}]{michoulier_2020}
Michoulier E.,  Toubin C.,  Simon A.,  Mascetti J.,  Aupetit C.,   Noble J.~A.,
   2020, \mn@doi [J. Phys. Chem. C] {10.1021/acs.jpcc.9b09499}, 124, 2994

\bibitem[\protect\citeauthoryear{Miliordos, Apr\'{a}  \& Xantheas}{Miliordos
  et~al.}{2016}]{miliordos_2016}
Miliordos E.,  Apr\'{a} E.,   Xantheas S.~S.,  2016, \mn@doi [J. Chem. Theory
  Comput.] {10.1021/acs.jctc.6b00668}, 12, 4004

\bibitem[\protect\citeauthoryear{Mukhopadhyay, Cole  \& Saykally}{Mukhopadhyay
  et~al.}{2015}]{MUKHOPADHYAY_2015}
Mukhopadhyay A.,  Cole W.~T.,   Saykally R.~J.,  2015, \mn@doi [Chem. Phys.
  Lett.] {https://doi.org/10.1016/j.cplett.2015.04.016}, 633, 13

\bibitem[\protect\citeauthoryear{Naumkin \& Knowles}{Naumkin \&
  Knowles}{1995}]{naumkin_1995}
Naumkin F.~Y.,  Knowles P.~J.,  1995, \mn@doi [J. Chem. Phys.]
  {10.1063/1.470224}, 103, 3392

\bibitem[\protect\citeauthoryear{Neese}{Neese}{2012}]{neese_2012}
Neese F.,  2012, \mn@doi [WIREs Comput. Mol. Sci.] {10.1002/wcms.81}, 2, 73

\bibitem[\protect\citeauthoryear{Neese \& Valeev}{Neese \&
  Valeev}{2011}]{neese_2011}
Neese F.,  Valeev E.~F.,  2011, \mn@doi [J. Chem. Theory Comput.]
  {10.1021/ct100396y}, 7, 33

\bibitem[\protect\citeauthoryear{Neese, Hansen  \& Liakos}{Neese
  et~al.}{2009}]{neese_2009}
Neese F.,  Hansen A.,   Liakos D.~G.,  2009, \mn@doi [J. Chem. Phys.]
  {10.1063/1.3173827}, 131, 064103

\bibitem[\protect\citeauthoryear{Noble, Michoulier, Aupetit  \& Mascetti}{Noble
  et~al.}{2020}]{noble_2020}
Noble J.,  Michoulier E.,  Aupetit C.,   Mascetti J.,  2020, \mn@doi [Astronomy
  \& Astrophysics] {10.1051/0004-6361/202038568}, 644, A22

\bibitem[\protect\citeauthoryear{Perdew, Burke  \& Ernzerhof}{Perdew
  et~al.}{1996}]{perdew_1996}
Perdew J.~P.,  Burke K.,   Ernzerhof M.,  1996, \mn@doi [Phys. Rev. Lett.]
  {10.1103/PhysRevLett.77.3865}, 77, 3865

\bibitem[\protect\citeauthoryear{Pl\'{i}va \& Johns}{Pl\'{i}va \&
  Johns}{1983}]{pliva_1983}
Pl\'{i}va J.,  Johns J. W.~C.,  1983, \mn@doi [Can. J. Phys.]
  {10.1139/p83-037}, 61, 269

\bibitem[\protect\citeauthoryear{Pl\'{i}va \& Johns}{Pl\'{i}va \&
  Johns}{1984}]{pliva_1984}
Pl\'{i}va J.,  Johns J.,  1984, \mn@doi [J. Mol. Spectrosc.]
  {10.1016/0022-2852(84)90011-0}, 107, 318

\bibitem[\protect\citeauthoryear{Pl\'{i}va \& Pine}{Pl\'{i}va \&
  Pine}{1982}]{pliva_1982}
Pl\'{i}va J.,  Pine A.,  1982, \mn@doi [J. Mol. Spectrosc.]
  {10.1016/0022-2852(82)90284-3}, 93, 209

\bibitem[\protect\citeauthoryear{Pl\'{i}va, Esherick  \& Owyoung}{Pl\'{i}va
  et~al.}{1987}]{pliva_1987}
Pl\'{i}va J.,  Esherick P.,   Owyoung A.,  1987, \mn@doi [J. Mol. Spectrosc.]
  {10.1016/0022-2852(87)90106-8}, 125, 393

\bibitem[\protect\citeauthoryear{Pl\'{i}va, Johns  \& Lu}{Pl\'{i}va
  et~al.}{1996}]{pliva_1996}
Pl\'{i}va J.,  Johns J.,   Lu Z.,  1996, \mn@doi [Mol. Phys.]
  {10.1080/00268979600100581}, 87, 859

\bibitem[\protect\citeauthoryear{Raza}{Raza}{2012}]{raza_2012}
Raza Z.,  2012, Proton ordering and reactivity of ice, \url
  {https://discovery.ucl.ac.uk/id/eprint/1369753/}

\bibitem[\protect\citeauthoryear{Reckien, Eggers  \& Bredow}{Reckien
  et~al.}{2014}]{reckien_2014}
Reckien W.,  Eggers M.,   Bredow T.,  2014, \mn@doi [Beilstein J. Org. Chem.]
  {10.3762/bjoc.10.185}, 10, 1775

\bibitem[\protect\citeauthoryear{Respondek \& Benoit}{Respondek \&
  Benoit}{2009}]{respondek_2009}
Respondek I.,  Benoit D.~M.,  2009, \mn@doi [J. Chem. Phys.]
  {10.1063/1.3193708}, 131, 054109

\bibitem[\protect\citeauthoryear{Salama}{Salama}{2008}]{salama_2008}
Salama F.,  2008, \mn@doi [Proc. {IAU}] {10.1017/S1743921308021960}, 4, 357

\bibitem[\protect\citeauthoryear{Salter, Stubbing, Brigham  \& Brown}{Salter
  et~al.}{2021}]{salter_2021}
Salter T.~L.,  Stubbing J.~W.,  Brigham L.,   Brown W.~A.,  2021, \mn@doi
  [Front. Astron. Space Sci.] {10.3389/fspas.2021.644277}, 8, 28

\bibitem[\protect\citeauthoryear{Salzmann}{Salzmann}{2019}]{salzmann_2019}
Salzmann C.~G.,  2019, \mn@doi [J. Chem. Phys.] {10.1063/1.5085163}, 150,
  060901

\bibitem[\protect\citeauthoryear{Schuhmann et~al.,}{Schuhmann
  et~al.}{2019}]{schuhmann_2019}
Schuhmann M.,  et~al., 2019, \mn@doi [Astron. Astrophys.]
  {10.1051/0004-6361/201834666}, 630, A31

\bibitem[\protect\citeauthoryear{Scribano \& Benoit}{Scribano \&
  Benoit}{2008}]{Scribano_2008}
Scribano Y.,  Benoit D.~M.,  2008, \mn@doi [Chem. Phys. Lett.]
  {10.1016/j.cplett.2008.05.001}, 458, 384

\bibitem[\protect\citeauthoryear{Scribano, Lauvergnat  \& Benoit}{Scribano
  et~al.}{2010}]{scribano_2010}
Scribano Y.,  Lauvergnat D.~M.,   Benoit D.~M.,  2010, \mn@doi [J. Chem. Phys.]
  {10.1063/1.3476468}, 133, 094103

\bibitem[\protect\citeauthoryear{Senent, Palmieri, Carter  \& Handy}{Senent
  et~al.}{2002}]{senent_2002}
Senent M.-L.,  Palmieri P.,  Carter S.,   Handy N.~C.,  2002, \mn@doi [Chem.
  Phys. Lett.] {10.1016/S0009-2614(01)01327-6}, 354, 1

\bibitem[\protect\citeauthoryear{Silva \& Devlin}{Silva \&
  Devlin}{1994}]{silva_1994}
Silva S.~C.,  Devlin J.~P.,  1994, \mn@doi [J. Phys. Chem.]
  {10.1021/j100093a027}, 98, 10847

\bibitem[\protect\citeauthoryear{Sivaraman, Mukherjee, Subramanian  \&
  Banerjee}{Sivaraman et~al.}{2014}]{sivaraman_2014}
Sivaraman B.,  Mukherjee R.,  Subramanian K.,   Banerjee S.,  2014, \mn@doi
  [Astrophys. J.] {10.1088/0004-637X/798/2/72}, 798, 72

\bibitem[\protect\citeauthoryear{Stephenson, Radloff  \& Rice}{Stephenson
  et~al.}{1984}]{stephenson_1984}
Stephenson T.~A.,  Radloff P.~L.,   Rice S.~A.,  1984, \mn@doi [J. Chem. Phys.]
  {10.1063/1.447800}, 81, 1060

\bibitem[\protect\citeauthoryear{Suzuki, Green, Bumgarner, Dasgupta, Goddard
  \& Blake}{Suzuki et~al.}{1992}]{suzuki_1992}
Suzuki S.,  Green P.~G.,  Bumgarner R.~E.,  Dasgupta S.,  Goddard W.~A.,
  Blake G.~A.,  1992, \mn@doi [Science] {10.1126/science.257.5072.942}, 257,
  942

\bibitem[\protect\citeauthoryear{Szczepaniak \& Person}{Szczepaniak \&
  Person}{1972}]{szczepaniak_1972}
Szczepaniak K.,  Person W.~B.,  1972, \mn@doi [Spectrochim. Acta A Mol. Biomol.
  Spectrosc.] {10.1016/0584-8539(72)80004-7}, 28, 15

\bibitem[\protect\citeauthoryear{Tabor, Kusaka, Walsh, Zwier  \&
  Sibert~III}{Tabor et~al.}{2015}]{tabor_2015}
Tabor D.~P.,  Kusaka R.,  Walsh P.~S.,  Zwier T.~S.,   Sibert~III E.~L.,  2015,
  \mn@doi [J. Phys. Chem. A] {10.1021/acs.jpca.5b06954}, 119, 9917

\bibitem[\protect\citeauthoryear{Thrower, Collings, Rutten  \&
  McCoustra}{Thrower et~al.}{2009}]{thrower_2009}
Thrower J.,  Collings M.,  Rutten F.,   McCoustra M.,  2009, \mn@doi [J. Chem.
  Phys.] {10.1063/1.3267634}, 131, 244711

\bibitem[\protect\citeauthoryear{Ulusoy, Scribano, Benoit, Tschetschetkin,
  Maurer, Koslowski  \& Ziemann}{Ulusoy et~al.}{2011}]{ulusoy_2011}
Ulusoy I.~S.,  Scribano Y.,  Benoit D.~M.,  Tschetschetkin A.,  Maurer N.,
  Koslowski B.,   Ziemann P.,  2011, \mn@doi [Phys. Chem. Chem. Phys.]
  {10.1039/C0CP01289K}, 13, 612

\bibitem[\protect\citeauthoryear{{VandeVondele} \& Hutter}{{VandeVondele} \&
  Hutter}{2007}]{vandevondele_2007}
{VandeVondele} J.,  Hutter J.,  2007, \mn@doi [J. Chem. Phys.]
  {10.1063/1.2770708}, 127, 114105

\bibitem[\protect\citeauthoryear{{VandeVondele}, Krack, Mohamed, Parrinello,
  Chassaing  \& Hutter}{{VandeVondele} et~al.}{2005}]{vandevondele_2005}
{VandeVondele} J.,  Krack M.,  Mohamed F.,  Parrinello M.,  Chassaing T.,
  Hutter J.,  2005, \mn@doi [Comput. Phys. Commun.]
  {10.1016/j.cpc.2004.12.014}, 167, 103

\bibitem[\protect\citeauthoryear{Visser, Geers, Dullemond, Augereau,
  Pontoppidan  \& van Dishoeck}{Visser et~al.}{2007}]{visser_2007}
Visser R.,  Geers V.,  Dullemond C.,  Augereau J.-C.,  Pontoppidan K.,   van
  Dishoeck E.,  2007, \mn@doi [Astron. Astrophys.]
  {10.1051/0004-6361:20066829}, 466, 229

\bibitem[\protect\citeauthoryear{Waite, Young, Cravens, Coates, Crary, Magee
  \& Westlake}{Waite et~al.}{2007}]{waite_2007}
Waite J.~H.,  Young D.~T.,  Cravens T.~E.,  Coates A.~J.,  Crary F.~J.,  Magee
  B.,   Westlake J.,  2007, \mn@doi [Science] {10.1126/science.1139727}, 316,
  870

\bibitem[\protect\citeauthoryear{Walsh, Millar, Nomura, Herbst, Weaver, Aikawa,
  Laas  \& Vasyunin}{Walsh et~al.}{2014}]{walsh_2014}
Walsh C.,  Millar T.~J.,  Nomura H.,  Herbst E.,  Weaver S.~W.,  Aikawa Y.,
  Laas J.~C.,   Vasyunin A.~I.,  2014, \mn@doi [Astron. Astrophys.]
  {10.1051/0004-6361/201322446}, 563, A33

\bibitem[\protect\citeauthoryear{Weigend}{Weigend}{2008}]{weigend_2008}
Weigend F.,  2008, \mn@doi [J. Comput. Chem.] {10.1002/jcc.20702}, 29, 167

\bibitem[\protect\citeauthoryear{Weigend \& H{\"a}ser}{Weigend \&
  H{\"a}ser}{1997}]{weigend_1997}
Weigend F.,  H{\"a}ser M.,  1997, \mn@doi [Theor. Chem. Acc. (Theoretica
  Chimica Acta)] {10.1007/s002140050269}, 97, 331

\bibitem[\protect\citeauthoryear{Wishnow, Gush  \& Ozier}{Wishnow
  et~al.}{1996}]{wishnow_1996}
Wishnow E.~H.,  Gush H.~P.,   Ozier I.,  1996, \mn@doi [J. Chem. Phys.]
  {10.1063/1.471056}, 104, 3511

\bibitem[\protect\citeauthoryear{W\'{o}jcik, G{\l}ug, Boczar  \&
  Boda}{W\'{o}jcik et~al.}{2014}]{wojcik_2014}
W\'{o}jcik M.~J.,  G{\l}ug M.,  Boczar M.,   Boda {\L}.,  2014, \mn@doi [Chem.
  Phys. Lett.] {10.1016/j.cplett.2014.08.018}, 612, 162

\bibitem[\protect\citeauthoryear{Woods \& Willacy}{Woods \&
  Willacy}{2007}]{woods_2007}
Woods P.~M.,  Willacy K.,  2007, \mn@doi [Astrophys. J. Lett.]
  {10.1086/511680}, 655, L49

\bibitem[\protect\citeauthoryear{Zamirri, Casassa, Rimola, Segado-Centellas,
  Ceccarelli  \& Ugliengo}{Zamirri et~al.}{2018}]{zamirri_2018}
Zamirri L.,  Casassa S.,  Rimola A.,  Segado-Centellas M.,  Ceccarelli C.,
  Ugliengo P.,  2018, \mn@doi [Mon. Not. R. Astron. Soc.]
  {10.1093/mnras/sty1927}, 480, 1427

\makeatother
\end{thebibliography}

\label{lastpage}

\newpage
\section*{Supplementary information}
\FloatBarrier
\begin{itemize}
    \item The geometry of the adsorbed benzene on our ice model is available as an XYZ file
    \item Our shifted version of the digitised spectrum from \citet{hagen_1983} is available as a CSV file 
\end{itemize}

\begin{table}
 \centering
\renewcommand\thetable{S1} 
\caption{Gas phase C--H stretching bands convergence with increasing number of allowed mode excitations (2-modes: SD, 3-modes:SDT and 4-modes:SDTQ). The SDTQ result is used as a reference and SD and SDT values are given as differences. All bands are intensity averaged.}
\begin{tabular}{ccccl}
\hline
\hline
& \multicolumn{3}{c}{VCIPSI [\icm]} &\\
 \cline{2-4}
Band & SD & SDT & SDTQ& Exp. \\
\hline
$\nu_{13}$& $-10$ & $+8$ & 3047& 3015 \citep{erlekam_2006}\\
$\nu_{7}$ & $-14$ & $+3$ & 3041&3057 \citep{hollinger_1978}\\
$\nu_{20}$& $+1$ & $0$ & 3059& 3048 \citep{pliva_1982} \\
$\nu_{2}$& $-8$ & $+2$ & 3081& 3074 \citep{jensen_1979}\\
\hline
\hline
\end{tabular}
\label{tbl:convergence}
\end{table}

\begin{landscape}
\begin{table}
 \centering
 \renewcommand\thetable{S2} 
 \caption{Vibrational frequencies of isolated benzene. Experimental data references for each line are given in the table. $^{\dag}$ = \citet{stephenson_1984}; $^{\ddag}$ line from \citet{erlekam_2006} }
\begin{tabular}{@{} ld{4.2}ld{4.0}d{2.6}d{4.0}lllllll @{}}
\hline
\hline
 Normal & \multicolumn{1}{c}{\textrm{Harmonic}} & Band label$^{\dag}$ & \multicolumn{1}{c}{PBE} & \multicolumn{1}{c}{\textrm{PBE-D3 diagonal}} & \multicolumn{1}{c}{\textrm{Hybrid}} & Exp & Wilson (Herzberg)& Reference & IR & Raman&$\mathcal{D}_{6h}$\\
 mode & \multicolumn{1}{c}{\textrm{(PBE-D3)}} && \multicolumn{1}{c}{VCIPSI4} &\multicolumn{1}{c}{\textrm{IR intensities}}& \multicolumn{1}{c}{\textrm{VCIPSI4}}& & label \\
\hline
1 & 396 & 16a& 397 & 1.33E-05 & 410 & 398 & 16(20) comb & \citep{brodersen_1956, brodersen_1959} & \xmark & \xmark & E$_{\rm 2u}$ \\
2 & 397 & 16b& 396 & 5.83E-05 & 410 & 398 & 16(20) comb & \citep{brodersen_1956, brodersen_1959} & \xmark & \xmark & E$_{\rm 2u}$ \\ 
3 & 603 & ``6b''& 602 & 6.70E-08 & 611 & 608 & 6(18) & \citep{hollinger_1978} & \xmark & \cmark & E$_{\rm 1g}$ \\
4 & 603 & ``6a''& 600 & 2.25E-07 & 609 & 608 & 6(18) & \citep{hollinger_1978} & \xmark & \cmark & E$_{\rm 1g}$ \\
5 & 663 & 11 & 694 & 16.61 & 718 & 674 & 11(4) & \citep{hollenstein_1990}& \cmark & \xmark & A$_{\rm 2u}$ \\
6 & 702 & 4 & 692 & 3.20E-03 & 706 & 707 & 4(8) comb & \citep{brodersen_1956, brodersen_1959} & \xmark & \xmark & B$_{\rm 1g}$ \\
7 & 833 & 10a& 848 & 2.30E-07 & 880 & 847 & 10(11) diff & \citep{pliva_1996}& \xmark & \cmark & E$_{\rm 1g}$ \\
8 & 834 & 10b& 846 & 4.51E-06 & 879 & 847 & 10(11) diff & \citep{pliva_1996}& \xmark & \cmark & E$_{\rm 1g}$ \\
9 & 948 & 17a& 954 & 1.84E-05 & 995 & 967 & 17(19) comb & \citep{brodersen_1956, brodersen_1959} & \xmark & \xmark & E$_{\rm 2u}$ \\
10 & 948 & 17b& 952 & 5.77E-05 & 994 & 967 & 17(19) comb & \citep{brodersen_1956, brodersen_1959} & \xmark & \xmark & E$_{\rm 2u}$ \\
11 & 976 & 5 & 983 & 1.37E-04 & 1022 & 990 & 5(7) comb & \citep{brodersen_1956, brodersen_1959} & \xmark & \xmark & B$_{\rm 1g}$ \\
12 & 991 & 1 & 984 & 3.84E-06 & 1000 & 993 & 1(2)& \citep{jensen_1979} & \xmark & \cmark & A$_{\rm 1g}$ \\
13 & 995 & 12 & 984 & 3.51E-03 & 1016 & 1010 & 12(6) & \citep{brodersen_1956, brodersen_1959} & \xmark & \xmark & B$_{\rm 1u}$ \\
14 & 1029 & 19b$\rightarrow$18b & 1025 & 0.9996& 1050 & 1038 & 18(14) & \citep{pliva_1984}& \cmark & \xmark & E$_{\rm 1u}$ \\
15 & 1038 & ``19a''$\rightarrow$18a & 1027 & 0.9788& 1052 & 1038 & 18(14) & \citep{pliva_1984}& \cmark & \xmark & E$_{\rm 1u}$ \\
16 & 1143 & 14$\rightarrow$15 & 1142 & 3.25E-04 & 1168 & 1148 & 15(10) diff & \citep{pliva_1996}& \xmark & \xmark & B$_{\rm 2u}$ \\
17 & 1158 & 8b$\rightarrow$9b & 1154 & 6.69E-07 & 1184 & 1178 & 9(17) & \citep{hollinger_1978} & \xmark & \cmark & E$_{\rm 1g}$ \\
18 & 1169 & 8a$\rightarrow$9a & 1157 & 2.59E-08 & 1186 & 1178 & 9(17) & \citep{hollinger_1978} & \xmark & \cmark & E$_{\rm 1g}$ \\
19 & 1333 & 3 & 1316 & 2.70E-07 & 1363 & 1350 & 3(3) comb & \citep{brodersen_1956, brodersen_1959} & \xmark & \xmark & A$_{\rm 1g}$ \\
20 & 1347 & 15$\rightarrow$14 & 1326 & 1.36E-03 & 1315 & 1309 & 14(9) & \citep{goodman-1989} & \xmark & \xmark & B$_{\rm 2u}$ \\
21 & 1463 & 18b$\rightarrow$19b & 1443 & 1.051 & 1489 & 1484 & 19(13) & \citep{pliva_1983}& \cmark & \xmark & E$_{\rm 1u}$ \\
22 & 1466 & 18a$\rightarrow$19a & 1444 & 1.086 & 1490 & 1484 & 19(13) & \citep{pliva_1983}& \cmark & \xmark & E$_{\rm 1u}$ \\
23 & 1591 & 9a$\rightarrow$8a & 1566 & 1.99E-07 & 1620 & 1601 & 8(16) fermi depert & \citep{pliva_1987}& \xmark & \cmark & E$_{\rm 1g}$ \\
24 & 1592 & 9b$\rightarrow$8b & 1562 & 1.58E-08 & 1616 & 1601 & 8(16) fermi depert & \citep{pliva_1987}& \xmark & \cmark & E$_{\rm 1g}$ \\
25 & 3080 & 13 & 2968 & 2.61E-02 & 3047 & 3015$^{\ddag}$ & 13(5) & \citep{erlekam_2006} & \xmark & \xmark & B$_{\rm 1u}$ \\
26 & 3088 & 7b & 2948 & 2.88E-05 & 3052 & 3057 & 7(15) & \citep{hollinger_1978} & \xmark & \cmark & E$_{\rm 1g}$ \\
27 & 3092 & 7a & 2935 & 2.74E-04 & 3040 & 3057 & 7(15) & \citep{hollinger_1978} & \xmark & \cmark & E$_{\rm 1g}$ \\
28 & 3104 & 20b& 2957 & 2.954 & 3071 & 3047.908 & 20(12) fermi res. & \citep{pliva_1982}& \cmark & \xmark & E$_{\rm 1u}$ \\
29 & 3107 & 20a& 2941 & 3.02 & 3047 & 3047.908 & 20(12) fermi res. & \citep{pliva_1982}& \cmark & \xmark & E$_{\rm 1u}$ \\
30 & 3115 & 2 & 2982 & 3.19E-04 & 3081 & 3074 & 2(1)& \citep{jensen_1979} & \xmark & \cmark & A$_{\rm 1g}$ \\
\hline
\hline
 \end{tabular}
 \label{tab:booktabs-gas}
\end{table}
\end{landscape}

\begin{table}
 \centering
 \renewcommand\thetable{S3} 
 \caption{Vibrational frequencies of benzene on ice. Hagen = \citet{hagen_1983}; Dawes = \citet{dawes_2018}; Szcz. = \citet{szczepaniak_1972}. $\star$=($\nu_{20}$ in original reference), $\dagger$=($\nu_{18}$ in original reference), $?$ indicates an ambiguous assignment and $\ddag$ corresponds to a band that we have re-attributed based on \citet{hagen_1983}'s observations.}
\begin{tabular}{@{} ld{4.2}ld{4.0}d{2.6}d{4.0}llllllllll @{}}
\hline
\hline
 Normal & \multicolumn{1}{c}{\textrm{Harmonic}} & Band label$^{\dag}$ & \multicolumn{1}{c}{PBE} & \multicolumn{1}{c}{\textrm{PBE-D3 diagonal}} & \multicolumn{1}{c}{\textrm{Hybrid}} & \multicolumn{3}{c}{Exp}& $\mathcal{D}_{6h}$\\
 \cline{7-9}
 mode & \multicolumn{1}{c}{\textrm{(PBE-D3)}} && \multicolumn{1}{c}{VCIPSI4} &\multicolumn{1}{c}{\textrm{IR intensities}}& \multicolumn{1}{c}{\textrm{VCIPSI4}}&Hagen & Dawes&Szcz.\\
\hline
1 & 331 & 16b & 339 & 33.87 & 355 & ---          & --- & ---          & E$_{\rm 2u}$ \\
2 & 411 & 16a & 405 & 0.2928 & 417 & ---          & --- & ---          & E$_{\rm 2u}$ \\
3 & 573 & 6a & 573 & 72.76 & 585 & ---          & --- & 601          & E$_{\rm 1g}$ \\
4 & 596 & 6b & 582 & 2.811 & 589 & ---          & --- & 601          & E$_{\rm 1g}$ \\
5 & 619 & 11 & 657 & 185.6 & 677 & 681          & --- & 704          & A$_{\rm 2u}$ \\
6 & 663 & 4  & 650 & 42.87 & 665 & ---          & --- & ---           & B$_{\rm 1g}$ \\
7 & 713 & 10b & 732 & 98.81 & 773 & 860          & 858 & 866          & E$_{\rm 1g}$ \\
8 & 773 & 10a & 783 & 42.47 & 816 & 860          & 858 & 866          & E$_{\rm 1g}$ \\
9 & 786 & 17b & 797 & 221.2 & 842 & 977          & 989 & 981          & E$_{\rm 2u}$ \\
10 & 907 & 5  & 906 & 84.93 & 945 & ---          & --- & 1004          & B$_{\rm 1g}$ \\
11 & 913 & 17a & 904 & 3.062 & 947 & 977          & 989 & 981          & E$_{\rm 2u}$ \\
12 & 946 & 1  & 935 & 327.4 & 951 & ---          & --- & 986          & A$_{\rm 1g}$ \\
13 & 981 & 18b & 971 & 96.79 & 994 & 1035$\star$ & 1035 & 1033$\star$ & E$_{\rm 1u}$ \\
14 & 991 & 12 & 975 & 13.33 & 1005 & 1012          & 1011 & 1011          & B$_{\rm 1u}$ \\
15 & 1028 & 18a & 1009 & 21.79 & 1034 & 1035$\star$ & 1035 & 1033$\star$& E$_{\rm 1u}$ \\
16 & 1066 & 9a & 1046 & 3.531 & 1078 & 1176          & 1177 & 1175          & E$_{\rm 1g}$ \\
17 & 1133 & 15 & 1123 & 7.236 & 1144 & 1148          & 1148 & 1148          & B$_{\rm 2u}$ \\
18 & 1156 & 9b & 1124 & 84.70 & 1152 & 1176          & 1177 & 1175          & E$_{\rm 1g}$ \\
19 & 1320 & 3  & 1292 & 19.22 & 1338 & ---          & --- & 1347          & A$_{\rm 1g}$ \\
20 & 1353 & 14 & 1329 & 45.80 & 1328 & 1305          & 1312 & 1310          & B$_{\rm 2u}$ \\
21 & 1399 & 8b & 1370 & 27.60 & 1427 & 1585          & 1586 & 1579          & E$_{\rm 1g}$ \\
22 & 1432 & 19b & 1404 & 117.4 & 1451 & 1478          & 1478 & 1475          & E$_{\rm 1u}$ \\
23 & 1455 & 19a & 1420 & 150.6 & 1462 & 1478          & 1478 & 1475          & E$_{\rm 1u}$ \\
24 & 1558 & 8a & 1423 & 349.9 & 1476 & 1585          & 1586 & 1579          & E$_{\rm 1g}$ \\
25 & 3058 & 13 & 2861 & 16.02 & 2978 & 3069?         & 3006 & 3063          & B$_{\rm 1u}$ \\
26 & 3066 & 7b & 2862 & 16.42 & 2975 & 3050          & --- & 3046          & E$_{\rm 1g}$ \\
27 & 3071 & 7a & 2959 & 79.11 & 3033 & 3050          & --- & 3046          & E$_{\rm 1g}$ \\
28 & 3082 & 20a & 2939 & 204.8 & 3004 & 3034$\dagger$ & 3034 & 3030$\dagger$ & E$_{\rm 1u}$ \\
29 & 3088 & 20b & 2892 & 118.0 & 3049 & 3034$\dagger$ & 3034 & 3030$\dagger$ & E$_{\rm 1u}$ \\
30 & 3096 & 2  & 2925 & 70.23 & 3031 & 3069?         & 3070\ddag & 3051          & A$_{\rm 1g}$ \\
\hline
\hline
 \end{tabular}
 \label{tab:booktabs-ice}
\end{table}

\end{document}